\documentclass[showpacs,amsmath,amsfonts,amssymb,aps,pra,superscriptaddress,10pt]{revtex4-2}
\usepackage{mathtools,amsthm,bm}
\usepackage[T1]{fontenc}
\usepackage{microtype}
\usepackage{xcolor}
\usepackage{graphicx}
\usepackage{booktabs}
\usepackage{tikz}
\usetikzlibrary{arrows.meta,calc,fit,positioning,shapes.geometric}
\usepackage[colorlinks=true,allcolors=blue!55!black]{hyperref}
\hypersetup{
  pdftitle={Perfect Wave Transfer in Nonconformal Tomonaga--Luttinger Liquids},
  pdfauthor={Dr. Davide Batic and Dr. Denys Dutykh},
  pdfsubject={Mathematics Department, Khalifa University of Science and Technology, Abu Dhabi, UAE.},
  pdfkeywords={perfect wave transfer, Tomonaga--Luttinger liquid, inverse spectral theory, compact boson, Darboux--Crum transformation}
}
\graphicspath{{figures/}{codes/generated/}}
\newtheorem{definition}{Definition}
\newtheorem{proposition}{Proposition}
\newtheorem{theorem}{Theorem}
\newtheorem{corollary}{Corollary}
\newtheorem{lemma}{Lemma}
\theoremstyle{remark}
\newtheorem{remark}{Remark}
\newtheorem{example}{Example}

\newcommand{\Hil}{\mathcal H}
\newcommand{\Fosc}{\mathcal F_{\mathrm{osc}}}

\newcommand{\Aosc}{\mathfrak A_{\mathrm{osc}}}
\newcommand{\Bop}{\mathcal B}
\newcommand{\id}{\mathbf 1}

\newcommand{\WoscAlg}{\mathfrak W_{\mathrm{osc}}}
\newcommand{\WzmAlg}{\mathfrak W_{\mathrm{zm}}}
\newcommand{\WfullAlg}{\mathfrak W_{\mathrm{full}}}
\newcommand{\WneutralAlg}{\mathfrak W_{\mathrm{neutral}}}

\begin{document}

\title{Perfect Wave Transfer in Nonconformal Tomonaga--Luttinger Liquids}

\author{Davide Batic}
\email{davide.batic@ku.ac.ae}
\thanks{Corresponding author.}
\affiliation{Mathematics Department, Khalifa University of Science and Technology, PO Box 127788, Abu Dhabi, United Arab Emirates}

\author{Denys Dutykh}
\email{denys.dutykh@ku.ac.ae}
\affiliation{Mathematics Department, Khalifa University of Science and Technology, PO Box 127788, Abu Dhabi, United Arab Emirates}

\date{\today}

\begin{abstract}
Perfect wave transfer (PWT) reproduces spatial reflection at a common finite time under a fixed Hamiltonian. We classify this process for open compact bosons and inhomogeneous Tomonaga--Luttinger liquids, distinguishing transfer of oscillator and neutral observables from transfer on the charged field algebra. In the single-charge model, the first two notions coincide. Global transfer additionally requires rephasing of the compact zero mode. For regular reflection-symmetric channels, exact oscillator transfer forces an odd multiple of the optical crossing time and a strictly ordered, parity-compatible spectrum differing from the conformal ladder at only
finitely many levels.  This spectral condition is also sufficient when the Robin boundary data induced by the physical profile are included.  Requiring Neumann conditions after the Liouville transformation instead forces conformality.  We construct an explicit, regular one-defect family for every $r>1/2$. Its transfer parameters are
precisely the reduced fractions $r=p/q$ with odd numerator and denominator, with shortest transfer time $pL/v_0$.  At the delay $3L/v_0$, the $r=3/5$ profile has numerical contrast $K_{\max}/K_{\min}\simeq 12.913$, compared with $81$ for $r=3$. Two independent conservative discretizations show second-order convergence on fixed mode bands. Occupation- and charge-constrained error estimates quantify robustness, while an exact homogeneous-chain calculation explains the $J^{2/3}$ resolved-band scale also observed in the engineered channels. These results provide a spectral
design method and finite-device benchmarks for quadratic quantum channels. An interacting microscopic realization remains open.
\end{abstract}

\maketitle

\section{Introduction}

The transfer of quantum information through a passive medium is a basic
problem in quantum communication. In finite spin chains and networks,
perfect state transfer (PST) occurs when a fixed Hamiltonian implements a spatial permutation, often a mirror inversion, at a prescribed time without local time-dependent control~\cite{Bose2003,ChristandlEtAl2004,
AlbaneseEtAl2004,ChristandlEtAl2005,NikolopoulosJex2014}. Related rephasing and mirror revivals occur in continuous one-particle quantum
mechanics~\cite{AronsteinStroud1997,Robinett2004}. For a quantum field, a common transfer time must align infinitely many normal modes. A compact field also has zero modes and distinct charge sectors, so the meaning of perfect transfer depends on which observables and coherent superpositions are included. One-dimensional critical systems provide a natural setting for this question. Their low-energy dynamics is often described by a compact boson or Tomonaga--Luttinger liquid (TLL), with propagation velocity $v$ and Luttinger parameter $K$~\cite{Cazalilla2004,CazalillaEtAl2011}. Both may vary with position. When $K$ is constant, the theory is an inhomogeneous conformal field theory (CFT). Moreover, the velocity variation can be absorbed into an optical coordinate, and correlations follow curved light-cone trajectories~\cite{DubailEtAl2017,DubailStephanCalabrese2017,Moosavi2024}. Such propagation fronts have been observed in a quantum field
simulator~\cite{TajikEtAl2023}. A varying $K$ couples the chiral sectors and produces nonconformal dynamics, with normal modes governed by a weighted Sturm--Liouville problem~\cite{GluzaMoosaviSotiriadis2022,Zettl2005}. Interfaces and endpoints have long played an essential role in open-boundary
bosonization and transport through inhomogeneous quantum wires~\cite{FabrizioGogolin1995,vonDelftSchoeller1998,MaslovStone1995,
SafiSchulz1995,Ponomarenko1995}. Here, the endpoint conditions also determine which spectra can support exact transfer.

A recent study formulated PWT for continuous inhomogeneous quantum fields, established a reflection-symmetry criterion for one-particle transfer in an inhomogeneous CFT, and related the general TLL problem to inverse Sturm--Liouville theory~\cite{MoosaviEtAl2026}. Several questions remain when transfer is required at the level of the
compact quantum field. The zero mode can prevent a single reflection
identity across charge sectors even when every oscillator rephases.
Weyl asymptotics fixes the high-energy spectral slope but does not alone fix the transfer time. Moreover, physical zero-current conditions become Robin conditions under the Liouville transformation in general. Imposing Neumann conditions on the transformed problem restricts the admissible profiles. Finally, a continuum transfer identity must be related to the finite bandwidth and dispersive corrections of a microscopic approximation~\cite{ImambekovGlazman2009,ImambekovSchmidtGlazman2012}.

We resolve these questions by combining the observable-algebra formulation with a necessary-and-sufficient spectral classification and explicit profile construction. The resulting nonconformal channels can transfer at odd multiples of the optical crossing time. A concrete comparison is shown in Fig.~\ref{Rdc}. The parameters $r=3$ and
$r=3/5$ give the same shortest transfer time $3L/v_0$, but their Luttinger profile contrasts are $81$ and approximately $12.913$, respectively. The rational family thus permits substantially smaller profile variation than its integer subfamily at the same delay. The classification has two parts. First, we formulate oscillator,
neutral-observable, and global transfer on bounded Weyl algebras. The first two notions coincide in the single-charge model studied here, because the fixed-charge zero-mode sectors are one-dimensional. Under the canonical scalar lift of reflection, global transfer additionally requires the quadratic charge-lattice condition $T/(4W)\in\mathbb Z$.  A noncanonical lift can change this condition, as an explicit example shows. Second, exact oscillator phase matching and regular fixed-Robin
Sturm--Liouville asymptotics force the transfer time to be an odd multiple of $L/v_0$ and the spectrum to coincide with the conformal ladder above finitely many levels. At the geometric time $L/v_0$, no spectral defect is possible. Moreover, for massless TLLs with Liouville--Neumann endpoints, the Miura factorization and heat invariants imply conformal rigidity. With the induced symmetric Robin data included, Hald's inverse theorem gives the converse classification. Every strictly ordered parity-compatible finite-defect sequence has a unique even potential and Robin parameter. Its positive zero mode reconstructs the TLL profile, up to scale and optical
reparametrization.

A Darboux--Crum transformation yields a complete explicit one-defect family for $r>1/2$, with dimensionless spectrum $0,r^{-2},4,9,\ldots$.
Oscillator PWT occurs precisely for reduced fractions $r=p/q$ with odd
$p$ and $q$, and its shortest time is $pL/v_0$.  Furthermore, the integer family is a restricted subfamily and the rational branch $r=p/(p+2)$ approaches the homogeneous channel smoothly as odd $p\to\infty$, while its transfer delay increases. An identity relating the endpoint slope to the logarithmic profile gradient quantifies the required inhomogeneity. We then examine how these exact continuum results survive finite-band and finite-site approximations. Small phase errors are controlled on a one-particle band, but unrestricted bosonic occupation can amplify them. We also derive occupation- and charge-constrained errors and first variations under profile, endpoint, and reflection-breaking perturbations. Two independent conservative discretizations exhibit $J^{-2}$ convergence on
fixed bands. Moreover, an exactly solvable homogeneous chain explains the $J^{2/3}$ scale for the growing resolved band. The engineered-channel data follow the corresponding parameter-free prediction over the tested range. Notice that a uniform theorem at this growing-band scale is not established for the engineered channels. We observed that the $r=3/5$ design is less sensitive to the tested endpoint pinning than $r=3$, but is not uniformly more robust under bulk perturbations. The finite-site models provide a controlled benchmark for quadratic oscillator dynamics. However, the construction of interacting microscopic systems with a compact charge rotor is left for future work.

The paper is organized as follows. Section~\ref{pwt_notions} specifies the self-adjoint model, represented algebras, and reflection lift. In  Section~\ref{spectral_arithmetic}, we derive the spectral and transfer-time constraints. Section~\ref{heat_rigidity} establishes the endpoint-dependent classification and constructs the rational family. Robustness and finite-site approximations are treated in  Section~\ref{finite_band_robustness} while Section~\ref{conclusions} discusses the physical implications and open problems. Notice that supporting reconstruction, inverse-spectral, and perturbative arguments are collected in the appendices.

\section{Global, sectorwise, and oscillator perfect wave transfer}
\label{pwt_notions}

We formulate PWT for an open compact boson and
establish how the transfer condition depends on the chosen observable
algebra. Starting from the self-adjoint dynamics and the canonical scalar action of reflection, we show that oscillator and sectorwise transfer coincide in the single-charge model. However, global transfer requires an additional arithmetic condition that aligns the compact zero-mode phases across charge sectors. The conformal TLL provides an explicit illustration of how this condition determines which coherent encodings can be transferred perfectly.

\subsection{Self-adjoint dynamics and compactification}\label{sad_c}

We begin by specifying the self-adjoint realization of the TLL mode
operator. The spatially varying parameters $v$ and $K$ determine its
natural Hilbert-space weight and enter the subsequent construction of
the zero mode, the optical coordinate, and the transformed boundary
conditions. More precisely, we consider the system on the interval $I=[-L/2,L/2]$. Throughout, unless a stronger hypothesis is stated, we assume
\begin{equation}\label{sr}
v,K\in W^{1,\infty}(I),\qquad
0<v_-\leq v\leq v_+,\qquad 
0<K_-\leq K\leq K_+.
\end{equation}
Let us set $\mathsf a=vK$, $w=K/v$, and $\Hil_1=L^2(I,w\,dx)$.  We formulate the TLL mode problem through the closed nonnegative
sesquilinear form
\begin{equation}\label{closed_form}
\mathfrak a[u,z]=\int_I\mathsf a(x)u'(x)\overline{z'(x)}~dx
\end{equation}
with $u,z\in\operatorname{dom}\mathfrak a=H^1(I)$. The following proposition identifies the self-adjoint operator associated with this quadratic form and establishes the spectral properties needed for the transfer analysis.
\begin{proposition}\label{wto}
The form in \eqref{closed_form} is densely defined, closed, and
nonnegative. Its self-adjoint operator is
\begin{equation}\label{wtop}
Au=-\frac{1}{w(x)}\frac{d}{dx}\left(\mathsf a(x)u'(x)\right),
\end{equation}
with domain
\begin{equation}\label{wtdom}
\operatorname{dom}A=\left\{u\in H^1(I):\mathsf a u'\in H^1(I),~ 
(\mathsf a u')(-L/2)=(\mathsf a u')(L/2)=0\right\}.
\end{equation}
Moreover, it has compact resolvent, simple eigenvalues
$0=\lambda_0<\lambda_1<\cdots$, and
\begin{equation}
\ker A=\operatorname{span}\{u_0\},\qquad
u_0(x)=W^{-1/2},\qquad
W=\int_I w(x)~dx.
\end{equation}
If $v$ and $K$ are even, the real eigenfunctions may be indexed so that
$u_n(-x)=(-1)^nu_n(x)$.
\end{proposition}
The proof is given in Appendix~\ref{operator_asymptotics}. The preceding proposition provides the mode basis for quantizing the open
field. We consider the Neumann compact boson with one integer charge
lattice and no independent winding sector, whose Hilbert space is
$\Hil=\ell^2(\mathbb Z)\otimes\Fosc$. The factor $\ell^2(\mathbb Z)$ is spanned by states of definite integer total charge, and $\Fosc$ is the bosonic Fock space generated by the nonzero standing-wave modes. For coherent superpositions of different charges, transfer also depends on the relative phases acquired by the zero mode in each sector. The observable algebra determines whether these intersector coherences enter the transfer criterion.

\subsection{PWT on an observable algebra}

We formulate PWT by requiring the dynamical and reflection actions to agree on a specified algebra of bounded observables. For a compact boson, the choice of observables and admissible states determines whether the transfer criterion probes the relative phases generated by the zero-mode dynamics across charge sectors. We underline that all exact statements below concern bounded observables. The symbols $\phi(x)$, $\Pi(x)$, and vertex fields are operator-valued distributions and are used only as mode-expansion or quadratic-form notation. The rigorous transfer identities are stated for bounded Weyl operators. Let $P$ be a unitary involution implementing
reflection and set $P^2=\id$ and $\gamma_P(A):=P^\dagger A P$. With $U(t)=e^{-iHt}$, we write $\alpha_t(A):=U(t)^\dagger A U(t)$. For a represented $C^*$-algebra $\mathfrak A\subset\Bop(\Hil)$ of bounded observables, we denote its commutant in $\Bop(\Hil)$ by $\mathfrak A'$.
\begin{definition}
The dynamics exhibits PWT on $\mathfrak A$ at time $T>0$ if
\begin{equation}\label{algebraic_pwt}
U(T)^\dagger A U(T)=P^\dagger A P
\end{equation}
for every $A\in\mathfrak A$. Equivalently, every normal state $\omega$ on the represented von Neumann algebra $\mathfrak A^{\prime\prime}$ satisfies $\omega(\alpha_T(A))=\omega(\gamma_P(A))$. The expectation of $A$ at time $T$ equals the expectation of its reflected image at time zero.
\end{definition}
The transfer condition can be characterized by comparing the evolution
operator with reflection. The following proposition shows that their
relative unitary must commute with every observable in $\mathfrak A$.
\begin{proposition}
Relation \eqref{algebraic_pwt} holds if and only if $V(T):=U(T)P^\dagger\in\mathfrak A'$.
\end{proposition}

\begin{proof}
Notice that \eqref{algebraic_pwt} implies $V(T)^\dagger A V(T)=A$ for every $A\in\mathfrak A$, and hence $[V(T),A]=0$. The converse follows by reversing the argument.
\end{proof}

\subsection{Weyl algebras and charge sectors}

We now construct the observable algebras associated with the oscillator
and compact zero-mode degrees of freedom. Their commutants determine the corresponding transfer conditions. For the nonzero modes, let
$\mathcal S_{\mathrm{osc}}:=c_{00}(\mathbb N;\mathbb R^2)$ be the real
vector space of finitely supported sequences $\zeta=(q_n,\pi_n)_{n\geq 1}$, equipped with the symplectic form
\begin{equation}
\sigma(\zeta,\zeta'):=\sum_{n\geq 1}(q_n\pi_n'-\pi_nq_n').
\end{equation}
The oscillator Weyl algebra $\WoscAlg$ is the unital $C^*$-algebra
generated by unitary symbols $W_{\mathrm{osc}}(\zeta)$ satisfying
\begin{equation}
W_{\mathrm{osc}}(\zeta)W_{\mathrm{osc}}(\zeta')=e^{-i\sigma(\zeta,\zeta')/2}W_{\mathrm{osc}}(\zeta+\zeta'),\qquad
W_{\mathrm{osc}}(\zeta)^*=W_{\mathrm{osc}}(-\zeta),
\end{equation}
where, in the regular Fock representation, we have
\begin{equation}
W_{\mathrm{osc}}(\zeta)=\exp\left(
i\sum_{n\geq 1}(q_n\phi_n+\pi_n\Pi_n)
\right).
\end{equation}
Exponentiating the canonical mode operators in this way gives bounded
unitaries. Moreover, the regular Fock representation of the Weyl relations is irreducible~\cite{DerezinskiGerard2013}. For the compact zero mode, we use $\Hil_{\mathrm{zm}}=\ell^2(\mathbb Z)$ with charge basis $\{|N\rangle\}_{N\in\mathbb Z}$, where $Q|N\rangle=N|N\rangle$.
The bounded unitaries
\begin{equation}
Z(\theta):=e^{i\theta Q},\qquad
S|N\rangle:=|N+1\rangle,\qquad
\theta\in\mathbb R/2\pi\mathbb Z
\end{equation}
obey the compact Weyl relation
\begin{equation}\label{cwr}
Z(\theta)S=e^{i\theta}SZ(\theta).
\end{equation}
We write $\WzmAlg:=C^*(Z(\theta),S)$ and define the represented charged field algebra by $\WfullAlg:=\WzmAlg\otimes_{\min}\WoscAlg$. Its gauge-invariant observable algebra is
$\WneutralAlg:=\left\{A\in\WfullAlg~\middle|~Z(\theta)AZ(\theta)^*=A
\ \text{for all }\theta\right\}$, where zero-mode operators act as the identity on the oscillator factor. The embedded oscillator algebra is
$\Aosc:=\id_{\mathrm{zm}}\otimes\WoscAlg$. In the standard representation on $\Hil=\Hil_{\mathrm{zm}}\otimes\Fosc$, let $\Pi_N$ denote the orthogonal projection onto the charge sector
$\Hil_N=\mathbb C|N\rangle\otimes\Fosc$. The relevant von Neumann closures and commutants are
\begin{eqnarray}
\WfullAlg^{\prime\prime}&=&\Bop(\Hil),\qquad\WfullAlg'=\mathbb C\id,
\label{fwi}\\
\WneutralAlg^{\prime\prime}&=&\mathop{\bigoplus}^{\ell^\infty}_{N\in\mathbb Z}\Bop(\Hil_N),\qquad
(\WneutralAlg^{\prime\prime})'=\left\{\sum_{N\in\mathbb Z}c_N\Pi_N
~\middle|~(c_N)\in\ell^\infty\right\},\qquad
\Aosc'=\Bop(\Hil_{\mathrm{zm}})\otimes\id_{\mathrm{osc}}.\label{nvnc}
\end{eqnarray}
To see this, let us recall that an operator commuting with every $Z(\theta)$ preserves each eigenspace of $Q$ and is therefore block diagonal with respect to the charge-sector decomposition of $\Hil$. Within each block, irreducibility of the oscillator representation makes an operator commuting with the oscillator algebra scalar. Finally, commutation with $S$ then makes these scalars independent
of $N$. This proves \eqref{fwi}, and the same block argument gives \eqref{nvnc}. Let us choose the canonical scalar lift of reflection
\begin{equation}\label{cpf}
P=P_{\mathrm{zm}}\otimes P_{\mathrm{osc}},\qquad
P_{\mathrm{zm}}=\id,\qquad
P_{\mathrm{osc}}=\Gamma(\mathsf p),\qquad
P_{\mathrm{osc}}\Omega=\Omega.
\end{equation}
Its action on the bounded generators is
\begin{equation}
PZ(\theta)P^{-1}=Z(\theta),\qquad
PSP^{-1}=S,\qquad
PW_{\mathrm{osc}}(\zeta)P^{-1}=W_{\mathrm{osc}}(\mathsf p\zeta).
\end{equation}
Here, $(\mathsf p\zeta)_n=\varepsilon_n(q_n,\pi_n)$, where
$\varepsilon_n=\pm 1$ is the parity of the $n$th normal mode. For the quadratic dynamics considered below, evolution and reflection
preserve these algebras and act on them by $*$-automorphisms. Furthermore, point-field expressions are understood through the corresponding Weyl identities or their differentiated forms on a common invariant core. The three observable algebras give the following transfer conditions.
\begin{definition}
Let us assume $[H,Q]=[P,Q]=0$.
\begin{enumerate}
\item
Global PWT holds when $\alpha_T(A)=\gamma_P(A)$ for every $A\in\WfullAlg$. By \eqref{fwi} and the commutant characterization, this is equivalent to $U(T)=e^{i\chi}P$ on $\Hil$.
\item
Sectorwise PWT occurs when $\alpha_T(A)=\gamma_P(A)$ for every $A\in\WneutralAlg$. This condition can be written as $\Pi_NU(T)\Pi_N=
e^{i\chi_N}\Pi_NP\Pi_N$ for every $N\in\mathbb Z$, where the phase may depend on the charge sector.
\item
\emph{Oscillator PWT} is realized when $\alpha_T(A)=\gamma_P(A)$ for every $A\in\Aosc$. When the dynamics and reflection factorize into zero-mode and oscillator parts, as in \eqref{H_factorization} below, irreducibility of the Fock representation makes this equivalent to
\begin{equation}\label{osc_pwt}
U_{\mathrm{osc}}(T)=e^{i\chi_{\mathrm{osc}}}P_{\mathrm{osc}}
\end{equation}
on $\Fosc$. Notice that the zero-mode unitary is unrestricted by this condition.
\end{enumerate}
\end{definition}
These three notions of transfer are related by the inclusions between
their observable algebras. Global PWT implies sectorwise PWT, which in
turn implies oscillator PWT. In the present model, oscillator and sectorwise PWT are equivalent. Indeed, oscillator PWT places $U(T)P^\dagger$ in the commutant \eqref{nvnc}. Moreover, charge conservation makes its zero-mode factor diagonal in the simple spectrum of $Q$, placing it in $(\WneutralAlg^{\prime\prime})'$. Global PWT additionally requires the phases of different charge sectors to agree with a single overall phase. Clearly, more general charge-sector structures can lead to different transfer
conditions. For instance, a zero-mode space with charge multiplicities can separate oscillator and sectorwise transfer, while grouping charges into even and odd sectors can produce an intermediate transfer criterion. We next derive the compact zero-mode Hamiltonian and determine the phase-alignment condition for global PWT. The Hamiltonian and mode fields are expressed in quadratic-form notation, with transfer understood through their action on the bounded Weyl generators.

\subsection{Zero-mode dynamics and global transfer}

Let $I$ be the compact interval introduced at the beginning of Section~\ref{sad_c}. We consider the massless inhomogeneous TLL Hamiltonian on $I$ 
\begin{equation}\label{tll_hamiltonian}
H=\int_I\frac{dx}{2\pi}~v(x)\left[\frac{[\pi\Pi(x)]^2}{K(x)}+K(x)[\partial_x\phi(x)]^2\right]
\end{equation}
with $v(x),K(x)>0$, Neumann boundary conditions and compactification $\phi\sim\phi+2\pi$. Furthermore, we set
\begin{equation}\label{W_def}
w(x):=\frac{K(x)}{v(x)},\qquad
W:=\int_I w(x)~dx.
\end{equation}
Let $\{u_n\}_{n\geq 0}$ be the real orthonormal eigenbasis of
Proposition~\ref{wto}. On the finite-mode core, we have
\begin{equation}
\phi(x)=\sum_{n\geq 0}X_nu_n(x),\qquad
\Pi(x)=w(x)\sum_{n\geq 0}P_nu_n(x),\qquad
[X_n,P_m]=i\delta_{nm}.
\end{equation}
Substituting these mode expansions into the TLL Hamiltonian and using
orthonormality together with $\mathfrak a[u_n,u_m]=\lambda_n\delta_{nm}$ yields
\begin{equation}
H=\frac{\pi}{2}\sum_{n\geq 0}P_n^2+\frac{1}{2\pi}\sum_{n\geq 1}\lambda_nX_n^2.
\end{equation}
Since $u_0=W^{-1/2}$, the conserved charge and the zero-mode momentum
density are
\begin{equation}
Q:=\int_I\Pi(x)\,dx=\sqrt{W}P_0,
\qquad
\Pi_{\mathrm{zm}}(x)=\frac{w(x)}{W}Q.
\end{equation}
The corresponding field coordinate is obtained by projecting onto the
constant mode. At the level of the mode expansion, orthonormality gives
\begin{equation}
\bar\phi:=\frac{1}{W}\int_I w(x)\phi(x)~dx=\frac{X_0}{\sqrt W}.
\end{equation}
Compactification makes $\bar\phi$ an angular variable whose conjugacy to the integer charge $Q$ is encoded by the bounded unitaries
$S=e^{i\bar\phi}$ and $Z(\theta)=e^{i\theta Q}$ through the Weyl
relation~\eqref{cwr}. Here, $e^{i\bar\phi}$ denotes the charge-shift
operator, and $[\bar\phi,Q]=i$ serves as formal notation because the
discrete spectrum of $Q$ precludes a self-adjoint conjugate coordinate
satisfying the global Weyl relations. For the standard untwisted compactification, the charge eigenstates satisfy $Q|N\rangle=N|N\rangle$ and $S|N\rangle=|N+1\rangle$ with $N\in\mathbb Z$. Thus, all zero-mode statements below involve only $Q$, its bounded exponentials, and the bounded shift $S$. After isolating the state-independent vacuum energy, the Hamiltonian therefore factorizes exactly as
\begin{equation}\label{H_factorization}
H=E_{\mathrm{vac}}+H_{\mathrm{zm}}+H_{\mathrm{osc}},\qquad
H_{\mathrm{zm}}=\frac{\pi}{2W}Q^2,\qquad
H_{\mathrm{osc}}=\sum_{n\geq 1}E_n a_n^\dagger a_n.
\end{equation}
Thus, in a charge eigenstate $Q\lvert N\rangle=N\lvert N\rangle$, we have $E_N^{(0)}=\pi N^2/(2W)$, which is exact in the normalization of
\eqref{tll_hamiltonian}. For a homogeneous TLL, it reduces to the standard finite-size charging energy $\pi vN^2/(2KL)$ (see, for example, \cite{Cazalilla2004}). By choosing the canonical scalar lift in \eqref{cpf}, we specify that reflection fixes every charge state $|N\rangle$, thereby removing the freedom to assign charge-dependent parity signs allowed by $[P,Q]=0$. If the Sturm--Liouville operator is reflection symmetric, its nonzero eigenfunctions can be chosen with parity $\varepsilon_n=\pm 1$ and
\begin{equation}\label{osc_parity}
Pa_nP^{-1}=\varepsilon_n a_n.
\end{equation}
We recall that for a regular even Sturm--Liouville problem ordered by increasing eigenvalue, $\varepsilon_n=(-1)^n$. The following theorem concerns the induced automorphisms of $\Aosc$, $\WneutralAlg$ and $\WfullAlg$.

\begin{theorem}\label{zmc}
Let us assume \eqref{H_factorization}-\eqref{osc_parity}, the standard
charge lattice $\operatorname{spec}Q=\mathbb Z$ and the canonical
scalar lift specified in \eqref{cpf}. Then, the following statements hold.
\begin{enumerate}
\item 
For every $n\geq 1$, oscillator PWT occurs at time $T$ if and only if
\begin{equation}\label{osc_phase_condition}
e^{-iE_nT}=\varepsilon_n.
\end{equation}
With normal ordering, the oscillator phase in \eqref{osc_pwt} is fixed to one by the oscillator vacuum.
\item 
Whenever \eqref{osc_phase_condition} occurs, the full propagator is
\begin{equation}\label{twisted_parity}
U(T)=e^{-iE_{\mathrm{vac}}T}\exp\left(-\frac{i\pi T}{2W}Q^2\right)P.
\end{equation}
\item 
At every oscillator-PWT time, PWT holds in each fixed-charge sector  with
\begin{equation}\label{sector_phase}
\chi_N=-E_{\mathrm{vac}}T-\frac{\pi T}{2W}N^2
\end{equation}
modulo $2\pi$.
\item 
Global PWT is realized if and only if \eqref{osc_phase_condition} holds and
\begin{equation}\label{gac}
\frac{T}{4W}\in\mathbb Z.
\end{equation}
\end{enumerate}
\end{theorem}
\begin{proof}
For each nonzero mode, let $W_n(z):=\exp(z a_n^\dagger-\overline{z} a_n)$ with $z\in\mathbb C$. These bounded unitaries generate $\WoscAlg$. Their dynamical and parity automorphisms are
\begin{equation}
U_{\mathrm{osc}}(T)^\dagger W_n(z)U_{\mathrm{osc}}(T)=W_n(e^{iE_nT}z),\qquad
P_{\mathrm{osc}}^\dagger W_n(z)P_{\mathrm{osc}}=W_n(\varepsilon_nz).
\end{equation}
Equality of the two automorphisms for every $z$ and every $n$ is therefore equivalent to $e^{iE_nT}=\varepsilon_n$, i.e. \eqref{osc_phase_condition}. Since the Fock representation of $\WoscAlg$ is irreducible, the commutant characterization then gives
$U_{\mathrm{osc}}(T)=e^{i\chi_{\mathrm{osc}}}P_{\mathrm{osc}}$. Normal ordering and the oscillator vacuum fix $e^{i\chi_{\mathrm{osc}}}=1$.
If we combine this identity with the commuting factorization
\eqref{H_factorization}, we obtain \eqref{twisted_parity}. The relative unitary $U(T)P^\dagger$ is then a function of $Q$, so it belongs to the commutant of the neutral algebra and restricts to the scalar phase \eqref{sector_phase} on each $\Hil_N$. Global PWT on $\WfullAlg$ requires this relative unitary to lie in $\WfullAlg'=\mathbb C\id$. Since its value at $N=0$ is one after removing the vacuum phase, this is equivalent to $\exp\left(-i\pi TN^2/(2W)\right)=1$ for every $N\in\mathbb Z$. The $N=1$ condition is precisely $T/(4W)\in\mathbb Z$, and it is sufficient for every integer $N$.
\end{proof}
At this point, a remark is in order.
\begin{remark}
Let $P_{\mathrm{zm}}$ denote the operator implementing reflection on
the zero-mode Hilbert space $\Hil_{\mathrm{zm}}=\ell^2(\mathbb Z)$.
The canonical scalar lift specifies $P_{\mathrm{zm}}|N\rangle=|N\rangle$ for every $N\in\mathbb Z$.
This condition is stronger than commutation with $Q$, since a unitary
involution commuting with $Q$ can assign charge-dependent signs through
$P_{\mathrm{zm}}|N\rangle=\eta_N|N\rangle$, with $\eta_N=\pm 1$. For example, the choice $\widetilde P_{\mathrm{zm}}=Z(\pi)=(-1)^Q$
gives $\widetilde P_{\mathrm{zm}}|N\rangle=(-1)^N|N\rangle$. Combining this action with the oscillator reflection operator $P_{\mathrm{osc}}$ defines the unitary involution $\widetilde P=\widetilde P_{\mathrm{zm}}\otimes P_{\mathrm{osc}}$ on $\Hil$. This operator commutes with $Q$ and implements spatial reflection composed with an internal $\mathbb Z_2$ transformation, whose action on the zero-mode generators is
\begin{equation}
\widetilde P Z(\theta)\widetilde P^{-1}=Z(\theta),\qquad
\widetilde P S\widetilde P^{-1}=-S.
\end{equation}
At an oscillator-PWT time, global PWT relative to this lift therefore
requires
\begin{equation}
e^{-i\pi TN^2/(2W)}=(-1)^N
\end{equation}
with $N\in\mathbb Z$. Taking $N=1$ gives
\begin{equation}
\frac{T}{2W}\in2\mathbb Z+1,
\end{equation}
which also ensures the required equality for every integer $N$ because
$N^2\equiv N\pmod2$. The change from \eqref{gac} illustrates how the
global transfer condition depends on the chosen lift of reflection,
including its action on the charge-shift operator $S$.
\end{remark}
The extra unitary in \eqref{twisted_parity} is invisible on the neutral fixed-point algebra but is detected by bounded charge-changing observables. For $m\in\mathbb Z$, let us introduce the charge-$m$ spectral subspace 
\begin{equation}
\WfullAlg^{(m)}:=\left\{B\in\WfullAlg:Z(\theta)BZ(\theta)^*=e^{im\theta}B\text{ for every }\theta\right\}. 
\end{equation}
For example, $S^m\otimes W_{\mathrm{osc}}(\zeta)$ belongs to
$\WfullAlg^{(m)}$. Let us set $a=\pi T/(2W)$. For every bounded
$B_m\in\WfullAlg^{(m)}$, functional calculus for $Q$ gives
\begin{equation}
e^{iaQ^2}B_me^{-iaQ^2}=B_m~e^{ia(2mQ+m^2)}.
\end{equation}
Hence, at an oscillator-PWT time, we find
\begin{equation}\label{charged_twist}
U(T)^\dagger B_m U(T)=\gamma_P(B_m)e^{ia(2mQ+m^2)}.
\end{equation}
For $m=0$, the additional phase factor in \eqref{charged_twist}
equals the identity, so oscillator PWT gives exact transfer on
$\WneutralAlg$. Requiring the same cancellation for all integer charges
$m$ and in every charge sector yields the global condition \eqref{gac}. To connect this result with the usual bosonization notation, let us consider the vertex fields $F_m(x)$, which carry charge $m$ and provide distributional representatives of the same charge grading. Their transfer relations express the phase dependence in \eqref{charged_twist}, with appropriate smearing, renormalization,
and a specification of common operator domains. The formulation in
terms of bounded Weyl operators therefore provides a precise basis
for interpreting these field relations and distinguishing oscillator,
sectorwise, and global PWT. The same phase-matching argument applies to more general zero-mode charge lattices and choices of reflection, leading to the following extension.

\begin{remark}
Let the allowed zero-mode charges form a lattice $\Gamma$ and let
the zero-mode reflection operator act as $P_{\mathrm{zm}}|q\rangle=\eta_q|q\rangle$, where $|\eta_q|=1$
and $\eta_q^2=1$ for an involutive lift. At an oscillator-PWT time,
global transfer requires a single phase $\chi_0$ satisfying
\begin{equation}\label{glc}
\exp\left(-\frac{i\pi T}{2W}q^2\right)=e^{i\chi_0}\eta_q
\end{equation}
for every $q\in\Gamma$. For the canonical scalar lift, $\eta_q=1$. Choosing any reference charge $q_\star\in\Gamma$, \eqref{glc} is then equivalent to
\begin{equation}
\frac{T}{4W}\left(q^2-q_\star^2\right)\in\mathbb Z
\end{equation}
for every $q\in\Gamma$. A background charge or twisted compactification is included by replacing $q^2$ with $(q-q_0)^2$. If we compose spatial reflection with a charge-dependent diagonal
internal phase, we modify the lift $\eta_q$ and the resulting global
transfer criterion. Since this internal transformation acts trivially
on neutral observables, their transfer condition remains unchanged. The global transfer criterion therefore depends on both the chosen
lift of reflection and the field algebra under consideration.
\end{remark}

\subsection{Oscillator and global transfer in conformal TLLs}

Let us suppose $K(x)=K$ is constant while $v(x)>0$ is even. Moreover, let $v_0$ be the harmonic-mean velocity determined by $\int_I dx/v(x)=L/v_0$.  Then, we have $W=KL/v_0$, $E_n=\pi n v_0/L$ and $
\varepsilon_n=(-1)^n$. For $r\in\mathbb N_0$, the oscillator condition is satisfied exactly at $T_r=(2r+1)L/v_0$. At these times, Theorem~\ref{zmc} yields the following result.
\begin{corollary}
For the untwisted charge lattice $\mathbb Z$ and the canonical scalar parity lift, the conformal TLL has global PWT at $T_r$ if and only if
\begin{equation}\label{K_arithmetic}
\frac{2r+1}{4K}\in\mathbb Z.
\end{equation}
Thus, oscillator PWT and fixed-charge sectorwise PWT hold for every positive $K$, whereas global PWT across all charge sectors is exceptional.
\end{corollary}
\begin{proof}
For constant $K$, we have $W=KL/v_0$. At the oscillator-transfer time
$T_r=(2r+1)L/v_0$, the global condition~\eqref{gac} becomes $T_r/(4W)=(2r+1)/(4K)\in\mathbb Z$. The oscillator and sectorwise statements follow from the conformal mode spectrum and the zero-mode classification theorem.
\end{proof}
The arithmetic condition~\eqref{K_arithmetic} also determines which
values of $K$ admit global transfer and fixes the earliest transfer time. For rational $K=A/B$, with coprime positive integers $A$ and $B$, it takes the form $[(2r+1)B]/(4A)\in\mathbb Z$. Since $2r+1$ is odd, this requires $B$ to be divisible by four. As $A$ and $B$ are coprime, $A$ and $2r+1$ must be odd and an integer multiple of $A$, respectively. The smallest admissible choice is $2r+1=A$, giving
\begin{equation}\label{egt}
T_{\mathrm{glob}}=\frac{AL}{v_0}.
\end{equation}
Thus, a global transfer time exists precisely when $K$ is rational
and its reduced denominator is divisible by four. Let us denote the zero-mode phase angle in charge sector $N$ by
\begin{equation}
\chi_N^{(0)}(T):=-T E_N^{(0)}=-\frac{\pi T}{2W}N^2.
\end{equation}
The role of the charge-dependent phases can be seen directly by taking
$K=1$ at the first oscillator-transfer time $T=L/v_0$. This implies 
$W=L/v_0$, so the corresponding phase factor is
\begin{equation}
e^{i\chi_N^{(0)}(L/v_0)}=
\begin{cases}
+1, & N\ \text{even},\\
-i, & N\ \text{odd}.
\end{cases}
\end{equation}
Since the zero-mode phase is constant within each even- or odd-charge
block, PWT is realized separately on both blocks. A coherent
superposition spanning the two blocks acquires a relative phase of
$-i$ between its odd- and even-charge components, preventing global PWT
at this time. Whether this relative phase enters the transfer criterion
depends on the observable algebra used to encode the quantum information.

\section{Spectral conditions and quantization of the transfer time}
\label{spectral_arithmetic}

We now determine the spectral conditions for oscillator PWT and the
transfer times compatible with a regular reflection-symmetric channel.
Exact transfer requires each mode to acquire the phase prescribed by
its reflection parity. Combined with the high-energy asymptotics of
the regular Sturm--Liouville problem, this condition forces every
transfer time to be an odd multiple of the optical crossing time and
the spectrum to coincide with the conformal ladder above finitely
many levels. We also show that transfer at the optical crossing time
requires the entire spectrum to follow this ladder, and determine the
earliest transfer time from the greatest common divisor of the
phase integers. We begin with the regular massless Sturm--Liouville problem associated with the open inhomogeneous TLL, i.e.
\begin{equation}
A u_n=\lambda_n u_n,\qquad
0=\lambda_0<\lambda_1<\lambda_2<\ldots,\qquad
E_n:=\sqrt{\lambda_n}.
\end{equation}
Throughout this section, we also assume that the velocity and Luttinger
profiles satisfy $v(-x)=v(x)$ and $K(-x)=K(x)$, and that the boundary
conditions are invariant under $x\mapsto -x$. Notice that since the eigenvalues are simple, the eigenfunctions have definite parity. Moreover, the Sturm oscillation theorem then gives $u_n(-x)=(-1)^n u_n(x)$ (see \cite{Zettl2005}). The oscillator Hamiltonian is the nonzero-mode term in \eqref{H_factorization}, while reflection acts according to \eqref{osc_parity} with $\varepsilon_n=(-1)^n$. By Theorem~\ref{zmc}, after removal of the state-independent vacuum phase, oscillator PWT at $T>0$ is equivalent to
\begin{equation}\label{epi}
E_nT=\pi m_n
\end{equation}
for every $n\geq 1$, where $m_0=0$ and the integers $m_n$ have the same parity as $n$. The strict ordering of $E_n$ and the identity $m_n=TE_n/\pi$ make this integer sequence strictly increasing. The corresponding phase condition is
\begin{equation}\label{epp}
e^{-iE_nT}=(-1)^n
\end{equation}
for every $n\geq 1$.
Equation \eqref{epi} associates each transfer time
with an integer sequence whose parity is fixed by reflection. The high-energy spectrum constrains both the allowed values of $T$
and the eventual form of this sequence. Its common divisors then
determine whether the chosen time is the earliest at which transfer
takes place.

\subsection{Weyl asymptotics and the optical crossing time}

The velocity profile defines the optical crossing time
\begin{equation}
\tau:=\int_{-L/2}^{L/2}\sqrt{\frac{w(x)}{\mathsf a(x)}}~dx
=\int_{-L/2}^{L/2}\frac{dx}{v(x)}=\frac{L}{v_0},
\end{equation}
where $w=K/v$ and $\mathsf a=vK$. Weyl
asymptotics~\cite{AtkinsonMingarelli1987,Zettl2005}
relates this time scale to the leading growth of the mode energies, that is
\begin{equation}\label{we}
E_n=\frac{\pi n}{\tau}+o(n)=\frac{\pi v_0 n}{L}+o(n).
\end{equation}
At an oscillator-PWT time $T$, substitution into \eqref{epi} yields 
\begin{equation}\label{cwc}
m_n=cn+o(n),\qquad
c:=\frac{T}{\tau}=\frac{Tv_0}{L}.
\end{equation}
Thus, the ratio between the transfer time and the optical crossing
time is the asymptotic slope of the phase integers. Notice that the ordering and parity of these integers already constrain this
ratio. Moreover, each increment $m_{n+1}-m_n$ is a positive odd integer, so $m_0=0$ implies $m_n\geq n$ with $c\geq 1$. Oscillator PWT therefore requires $T\geq \tau$. Determining which ratios $c$ are admissible requires more information than the leading Weyl estimate provides. For example, the sequence $m_{2k}=4k$ and $
m_{2k+1}=4k+1$ for nonnegative integers $k$ is strictly increasing, has the required alternating parity and satisfies $m_n=2n+O(1)$. It is therefore compatible with \eqref{cwc} at $c=2$. This arithmetic example illustrates the freedom left by the
$o(n)$ remainder. Its realization as a regular TLL spectrum would
require additional spectral conditions. This distinction also clarifies the comparison with \cite{MoosaviEtAl2026}. More precisely, Equation (48) in the supplementary material accompanying \cite{MoosaviEtAl2026} establishes the leading asymptotics in \eqref{we}, while the derivation of Equation (59) in the same material uses $v_0=L/T$, thereby identifying the shortest common transfer time with the optical crossing time.
That identification requires a separate argument, since the
leading Weyl estimate gives \eqref{cwc} with $c\geq 1$. We therefore keep $c$ undetermined at this stage and use sharper
spectral asymptotics below to establish the allowed transfer times.

\subsection{Spectral regularity and transfer-time quantization}

To sharpen the Weyl estimate, we express the mode equation in
Liouville normal form and identify the transformed boundary
conditions. For the regular open TLL, the physical zero-current
condition gives Neumann conditions on the original mode $u$. Moreover, the boundary conditions on the transformed mode also depend on the endpoint derivatives of $K$. For $x\in I=[a,b]=[-L/2,L/2]$, we introduce the coordinate
\begin{equation}\label{lca}
y(x)=\int_0^x\frac{v_0}{v(s)}~ds.
\end{equation}
Since $v$ is positive and even, the normalization of $v_0$ gives
$y(a)=a$ and $y(b)=b$. Thus, $y$ maps $I$ onto itself. Let $x(y)$ denote the inverse map. We define $s(y):=\sqrt{K(x(y))}$ and $\psi(y):=s(y)u(x(y))$. Furthermore, we assume $s\in W^{2,\infty}(I)$ throughout this subsection and the subsequent inverse spectral analysis. Since $s$ is continuous and strictly positive on the compact interval $I$, it is bounded away from zero. Together with $s\in W^{2,\infty}(I)$, this gives $s''/s\in L^\infty(I)$. All profiles constructed below
satisfy this regularity condition. Finally, primes denote derivatives with respect to $y$ in the remainder
of this subsection. At the endpoints, the physical Neumann
conditions become
\begin{equation}
\left(\frac{\psi}{s}\right)'(a)=0=\left(\frac{\psi}{s}\right)'(b)
=0.
\end{equation}
Since $s$ is even, the transformed mode satisfies the symmetric Robin conditions
\begin{equation}\label{lsra}
\psi'(a)=\rho\psi(a),\qquad
\psi'(b)=-\rho\psi(b),\qquad
\rho:=\frac{s'(a)}{s(a)}=-\frac{s'(b)}{s(b)}.
\end{equation}
The transformed eigenvalue problem is
\begin{equation}
\mathcal L_{\widehat{q},\rho}\psi:=\left[-v_0^2\frac{d^2}{dy^2}+\widehat{q}(y)\right]\psi=\lambda\psi,\qquad
\widehat{q}(y)=v_0^2\frac{s''(y)}{s(y)},
\end{equation}
with endpoint conditions as in \eqref{lsra}. The Liouville-Neumann case corresponds to $\rho=0$. At regular endpoints, this is equivalent to $\partial_xK(\pm L/2)=0$, as noted in \cite{MoosaviEtAl2026}. Hence, the parameter $\rho$ is determined by the Luttinger profile. The following estimate applies to every fixed real value of $\rho$.
\begin{proposition}\label{rhea}
Let $\rho\in\mathbb R$ be fixed and $\widehat{q}\in L^\infty(I)$ be real. By $(\lambda_n)_{n\geq 0}$, we denote the eigenvalues of
$\mathcal{L}_{\widehat{q},\rho}$ with boundary conditions
\eqref{lsra}, ordered increasingly from the lowest eigenvalue and counting multiplicity. Then, 
\begin{equation}\label{rla}
\lambda_n=\left(\frac{\pi v_0 n}{L}\right)^2+O(1).
\end{equation}
In particular, $\lambda_n>0$ for all sufficiently large $n$ 
and the corresponding energies $E_n=\sqrt{\lambda_n}$ satisfy
\begin{equation}\label{sea}
E_n=\frac{\pi v_0 n}{L}+O(n^{-1}).
\end{equation}
The constants in the remainders may depend on $L$, $v_0$,
$\rho$, and $\lVert\widehat{q}\rVert_\infty$ and are
independent of $n$.
\end{proposition}
The proof is given in Appendix~\ref{operator_asymptotics}.
The essential improvement over the leading Weyl estimate is
that the difference $E_n-(\pi v_0 n/L)$ tends to zero.
When combined with the integer phase condition, this
convergence quantizes the transfer time and forces the
high-energy spectrum to become exactly linear. The next
theorem makes this conclusion precise.
\begin{theorem}\label{ttq}
Let us suppose that the regular reflection-symmetric open TLL
satisfies \eqref{sea} and exhibits oscillator PWT at time $T>0$. By $m_n$, we denote the strictly increasing phase integers in \eqref{epi} with $m_0=0$ and $m_n$ of the same parity as $n$. Then, 
\begin{equation}\label{c_odd}
c=\frac{Tv_0}{L}\in\{1,3,5,\ldots\}.
\end{equation}
Moreover, there exists $N_\ast$ such that for every
$n\geq N_\ast$
\begin{equation}\label{efe}
m_n=cn,\qquad
E_n=\frac{\pi v_0 n}{L}.
\end{equation}
Thus, every oscillator-PWT time is an odd multiple of
the optical crossing time, i.e.
\begin{equation}
T=c\frac{L}{v_0}
\end{equation}
with $c\in\{1,3,5,\ldots\}$, and departures from the exactly linear spectrum are confined to finitely many low modes.
\end{theorem}
\begin{proof}
First of all, \eqref{epi} and \eqref{sea} yield
\begin{equation}\label{mn_small_remainder}
m_n=cn+r_n,\qquad 
r_n=O(n^{-1}).
\end{equation}
Let us define $d_n:=m_{n+1}-m_n$. Since the phase integers are
strictly increasing and alternate in parity, every $d_n$
is a positive odd integer. At the same time, $d_n=c+r_{n+1}-r_n\longrightarrow c$. A convergent sequence of positive odd integers is eventually constant, so its limit $c$ is itself a positive odd integer. Consequently, $m_n-cn$ is an integer for every $n$. Finally, \eqref{mn_small_remainder} shows that these
integers tend to zero and hence, vanish for all sufficiently
large $n$. This proves \eqref{c_odd} and \eqref{efe}.
\end{proof}
Notice that the proof uses only $r_n\to 0$, so the same conclusion follows if the remainder in \eqref{sea} is replaced by $o(1)$. At the optical crossing time, the conclusion extends to the entire spectrum. In this case, any excess in an early phase increment would persist at all higher indices, contradicting the eventual equality $m_n=n$.
\begin{corollary}\label{gtfl}
Under the hypotheses of Theorem~\ref{ttq}, if oscillator
PWT occurs at $T=L/v_0$, then, for every $n\geq 0$ $m_n=n$ and 
$E_n=\pi v_0 n/L$.
\end{corollary}
\begin{proof}
For $c=1$, Theorem~\ref{ttq} gives $m_n=n$ for all sufficiently large $n$. Since $m_0=0$ and each increment $d_n=m_{n+1}-m_n$ is at least one, 
\begin{equation}
m_n-n=\sum_{j=0}^{n-1}(d_j-1)
\end{equation}
is nonnegative and nondecreasing. As this sequence is zero for all sufficiently large $n$, it must be zero for every $n\geq 0$. Hence, $m_n=n$ for all $n\geq 0$. The energy formula follows from \eqref{epi}.
\end{proof}

\subsection{Shortest transfer time and finite spectral defects}

The spectral criterion identifies times at which oscillator PWT
occurs. To determine the shortest such time, we must also account
for common factors among the phase integers. The following
proposition gives the complete set of oscillator-PWT times
for a fixed spectrum.
\begin{proposition}\label{minimality_gcd}
Suppose oscillator PWT occurs at time $T>0$ with phase
integers $m_n$ satisfying \eqref{epi} and having the
same parity as $n$. Let us define $g:=\gcd\{m_n:n\geq 1\}$.
Then, $g$ is odd and the shortest positive oscillator-PWT
time is
\begin{equation}\label{minimal_time_gcd}
T_{\min}=\frac{T}{g}.
\end{equation}
All positive oscillator-PWT times for this spectrum are
\begin{equation}\label{pwt_times_gcd}
t_j=(2j+1)\frac{T}{g}
\end{equation}
with $j\in\mathbb N_0$. In particular, $T$ is the shortest transfer time if and only if $g=1$. We call such a time primitive.
\end{proposition}
\begin{proof}
Since $m_1$ is odd, its divisor $g$ is also odd. At time $T/g$, the mode phases satisfy $E_n T/g=\pi m_n/g$. Division by the odd integer $g$ preserves parity, so $m_n/g$ has the same parity as $n$. Thus, $T/g$ and all its positive odd multiples are oscillator-PWT times. Conversely, let $t=\alpha T$ be any positive oscillator-PWT time. Then, $\alpha m_n$ is an integer for every $n$. The greatest common divisor of the full sequence equals that of some finite subset, so B\'ezout's identity gives $\alpha g=k$ for a positive integer $k$. Since $\alpha m_1=k(m_1/g)$ must be odd, $k$ is odd. Hence, every positive oscillator-PWT time is an odd
multiple of $T/g$, proving \eqref{minimal_time_gcd} and
\eqref{pwt_times_gcd}.
\end{proof}
For the regular channels considered in Theorem~\ref{ttq}, this greatest common divisor can be computed from finitely many low modes. Moreover, the same theorem gives $m_n=cn$ for all $n\geq N_\ast$ where $c$ is a positive odd integer. Since the indices in this tail are consecutive, the greatest common divisor of its
phase integers is $c$. The remaining phase integers have the same parity as $cn$, so we may write
\begin{equation}\label{fsd}
m_n=cn+2s_n
\end{equation}
with $s_n\in\mathbb Z$ and $s_n=0$ for every $n\geq N_\ast$. The integers $s_n$ describe the finite spectral defects, namely the deviations from the exactly linear spectrum. It follows that $g=\gcd\bigl(c,s_1,\ldots,s_{N_\ast-1}\bigr)$ because $c$ is odd and the factors of two do not affect the common divisors. Combining this expression with \eqref{minimal_time_gcd} yields
$T_{\min}=cL/(gv_0)$. Thus, the candidate time $T=cL/v_0$ is primitive precisely when $c$ and the finite defect integers have
greatest common divisor one. These results specify the spectral form required for oscillator PWT. Establishing that such spectra arise from regular TLL profiles requires an inverse spectral
construction. Theorem~\ref{chr} supplies this construction for every strictly increasing sequence of the form
\eqref{fsd} with $m_0=0$ and $c$ a positive odd integer, when the potential and the symmetric Robin parameter are determined together
from the profile.

The spectral constraints and the minimality criterion
therefore have complementary roles. Regularity forces
the high-energy spectrum to become exactly linear and
restricts every oscillator-PWT time to an odd multiple
of the optical crossing time. The finite low-energy
deviations determine which of these times is the
shortest. If the spectrum is linear at every level,
then $g=c$ and $T_{\min}=L/v_0$. A shortest transfer
time exceeding the optical crossing time consequently
requires nonzero spectral defects with their common
divisors fixing the delay through the ratio $c/g$.

\section{Boundary conditions, spectral rigidity, and delayed transfer}\label{heat_rigidity}

The preceding section confines departures from the linear spectrum
to finitely many low modes. We now determine when these spectral
deviations can arise from a positive massless TLL profile.
The transformed boundary conditions play a central role.
For Liouville-Neumann endpoints, the first heat invariant,
combined with the Miura factorization, forces every spectral
deviation to vanish and the Luttinger parameter to be constant.
For the symmetric Robin conditions induced by a general open
profile, we use Hald's inverse spectral theorem to realize
every admissible finite deviation by determining the potential
and the endpoint parameter together. These results identify
the endpoint restriction responsible for Neumann rigidity
and establish the existence of nonconformal profiles with
delayed oscillator PWT. Throughout this section, we work on
$I=[a,b]=[-L/2,L/2]$ and use the rescaled Liouville operator
\begin{equation}
\mathcal S_V=-\frac{d^2}{dy^2}+V(y),\qquad
V(y):=\frac{\widehat{q}(y)}{v_0^2}=\frac{s''(y)}{s(y)}.
\end{equation}
We recall that $s(y)=\sqrt{K(x(y))}$, primes denote derivatives
with respect to $y$ and its eigenvalues are $\Lambda_n=E_n^2/v_0^2$. In particular, we compare them with the free Neumann spectrum
\begin{equation}\label{fns}
\Lambda_n^{(0)}=\left(\frac{\pi n}{L}\right)^2
\end{equation}
with $n\in\mathbb N_0$.

\subsection{Finite spectral defects and Neumann heat invariants}

We first consider a real potential $V\in C^\infty(I)$
with Neumann boundary conditions at $a$ and $b$.
The heat trace relates the spectrum to integrals of the
potential and its endpoint values. For the operator
$\mathcal S_V$, the standard Neumann heat-kernel
coefficients give \cite{Vassilevich2003}
\begin{eqnarray}
\Theta_V^N(t)&:=&\operatorname{Tr}e^{-t\mathcal{S}_V}=\sum_{n=0}^\infty e^{-t\Lambda_n}=\frac{L}{\sqrt{4\pi t}}+\frac{1}{2}-\frac{t}{4}\left[V(a)+V(b)\right]-\frac{t^{1/2}}{\sqrt{4\pi}}\int_a^b V(y)~dy\nonumber\\
&&+\frac{t^{3/2}}{\sqrt{4\pi}}\left\{\frac{1}{2}\int_a^b V(y)^2~dy
+\frac{1}{2}\left[V'(b)-V'(a)\right]\right\}+O(t^2).\label{nhte}
\end{eqnarray}
Here, the expansion is understood as $t\downarrow 0$.  For a real bounded potential, the leading difference from the
free Neumann heat trace still satisfies
\begin{equation}
\Theta_V^N(t)-\Theta_0^N(t)=-\frac{t^{1/2}}{\sqrt{4\pi}}\int_a^b V(y)~dy+o(t^{1/2})
\end{equation}
as $t\downarrow 0$. Thus, the argument yielding $\int_I V(y)~dy=0$ also applies to the bounded Liouville potentials considered above. Notice that the assumption $V\in C^\infty(I)$ justifies the displayed terms of order $t$ and $t^{3/2}$ in \eqref{nhte},
which depend on the endpoint values $V(a)$ and $V(b)$ and the derivatives $V'(a)$ and $V'(b)$. If the spectrum differs from the free Neumann spectrum at only finitely many levels, the difference of the two heat traces is analytic at $t=0$. Hence, its Taylor expansion contains only integer powers of $t$, which
constrains the half-integer coefficients in \eqref{nhte}.
\begin{proposition}\label{htfd}
Let $V\in C^\infty(I)$ be real. Further suppose that the Neumann
spectrum of $\mathcal{S}_V$ agrees with the free Neumann spectrum for every $n\geq N_\ast$,
\begin{equation}\label{finite_tail_equality}
\Lambda_n=\Lambda_n^{(0)}.
\end{equation}
Then, we have
\begin{equation}\label{fhri}
\int_a^b V(y)~dy=0
\end{equation}
and
\begin{equation}\label{second_heat_rigidity_identity}
\int_a^b V(y)^2~dy+V'(b)-V'(a)=0.
\end{equation}
For an even potential, the second identity becomes
\begin{equation}\label{eshi}
\int_{-L/2}^{L/2}V(y)^2~dy+2V'(L/2)=0.
\end{equation}
\end{proposition}
\begin{proof}
By \eqref{finite_tail_equality}, the difference of the heat traces is the finite sum
\begin{equation}
\Theta_V^N(t)-\Theta_0^N(t)=\sum_{n=0}^{N_\ast-1}
\left(e^{-t\Lambda_n}-e^{-t\Lambda_n^{(0)}}\right).
\end{equation}
Each summand is analytic at $t=0$ and vanishes there. Hence, the difference has a Taylor expansion containing only positive integer powers of $t$. This implies that the coefficients of $t^{1/2}$ and $t^{3/2}$ in \eqref{nhte} must equal their free values, which are zero. This gives \eqref{fhri} and \eqref{second_heat_rigidity_identity}. If $V$ is even, then $V'(a)=-V'(b)$, yielding \eqref{eshi}.
\end{proof}
The second identity relates the squared potential to its endpoint derivatives. If reflection across the endpoints produces a continuously differentiable periodic extension, both derivatives vanish. This additional regularity then forces the potential
itself to vanish.
\begin{corollary}
Under the hypotheses of Proposition~\ref{htfd}, we suppose that reflection across the endpoints extends $V$ to a $C^1$ potential on the circle of circumference $2L$. Then, $V\equiv 0$ on $I$ and
$\Lambda_n=\Lambda_n^{(0)}$ for every $n\geq 0$.
\end{corollary}
\begin{proof}
Continuous differentiability across the reflection points requires $V'(a)=V'(b)=0$. As a consequence, \eqref{second_heat_rigidity_identity} gives $\int_I V(y)^2~dy=0$. Since $V$ is real and continuous, it vanishes throughout $I$.
\end{proof}
For an even potential, \eqref{eshi} also gives $V'(L/2)\leq 0$  with strict inequality whenever $V$ is nonzero. This sign differs from $V'(L/2)\geq 0$ inferred from Equations~(58)--(60)
in the supplementary material accompanying \cite{MoosaviEtAl2026}. Under the additional smooth-reflection assumption, both arguments lead to $V\equiv 0$. Our heat-trace proof establishes this
conclusion directly from the spectral tail and the endpoint regularity.

\subsection{Neumann rigidity for massless TLLs}

For a massless TLL, the relation $V=s''/s$ connects the
Liouville potential directly to the positive Luttinger
profile. The additional endpoint condition $\partial_xK(\pm L/2)=0$ gives
\begin{equation}\label{sne}
s'(a)=s'(b)=0,
\end{equation}
where primes denote derivatives with respect to $y$. Under these conditions, the integral of $V$ can be written as the integral of a nonnegative square. The first heat identity therefore suffices to establish rigidity without an assumption about smooth reflection
of the potential across the endpoints.
\begin{theorem}\label{massless_neumann_rigidity}
Let $s\in W^{2,\infty}(I)$ be strictly positive and satisfy \eqref{sne}. Let us set $V=s''/s$ and equip $\mathcal{S}_V$ with Neumann boundary conditions. If its spectrum agrees with \eqref{fns} for all sufficiently large $n$, then $s$ and the corresponding Luttinger parameter $K$ are constant and $V=0$ almost everywhere on $I$.
\end{theorem}
\begin{proof}
Since $V$ is bounded, the first heat-trace identity gives \eqref{fhri}. Let us define $r=s'/s$. Then, $r\in W^{1,\infty}(I)$ and $V=r'+r^2$ almost everywhere. The endpoint conditions give $r(a)=r(b)=0$, so
\begin{equation}
\int_a^b V(y)~dy=r(b)-r(a)+\int_a^b r(y)^2~dy=\int_a^b r(y)^2~dy.
\end{equation}
Hence, \eqref{fhri} implies $\int_I r(y)^2~dy=0$. Since $r$ is continuous, $r=0$ throughout $I$. Therefore, $s$ and the corresponding Luttinger parameter are constant and $V=0$
almost everywhere.
\end{proof}
The spectral condition in this theorem follows from
oscillator PWT by Theorem~\ref{ttq}. As a consequence, within this endpoint class, exact transfer requires a constant Luttinger parameter and fixes the shortest transfer time to the optical crossing time.
\begin{corollary}
Let us consider a regular reflection-symmetric massless TLL
with $s\in W^{2,\infty}(I)$ and $\partial_xK(\pm L/2)=0$. If it exhibits oscillator PWT, then $K$ is constant and for every
$n\geq 0$
\begin{equation}\label{cfsr}
E_n=\frac{\pi v_0 n}{L}.
\end{equation}
Its oscillator-PWT times are precisely the positive odd multiples of $L/v_0$ and its shortest oscillator-PWT time is $T_{\min}=L/v_0$.
\end{corollary}
\begin{proof}
Theorem~\ref{ttq} shows that oscillator PWT forces the spectrum to agree with the free spectrum for all sufficiently large mode indices. Hence, Theorem~\ref{massless_neumann_rigidity} gives
constant $K$ and vanishing $V$, establishing \eqref{cfsr}.
For this spectrum, the first-mode phase condition
requires $Tv_0/L$ to be a positive odd integer. Every such value also satisfies the phase conditions for all higher modes, giving the stated transfer times and their minimum.
\end{proof}

\subsection{Complete spectral realization with induced Robin conditions}

We now allow the symmetric Robin parameter to be determined
together with the Liouville potential. In this class, every
spectrum satisfying the arithmetic conditions of
Section~\ref{spectral_arithmetic} can be realized by a regular
massless TLL. The construction uses Hald's inverse spectral
theorem to recover the potential and endpoint parameter,
then obtains the Luttinger profile from the positive
zero-energy ground state.
\begin{theorem}\label{chr}
Let us fix a positive odd integer $c$ and a strictly increasing
sequence of integers $(m_n)_{n\geq 0}$. We further assume that
$m_n$ has the same parity as $n$ and that $0=m_0<m_1<m_2<\cdots$ with $m_n=cn$ for all $n\geq N_\ast$.
Let us define the target spectrum by
\begin{equation}\label{htsg}
\Lambda_n=\left(\frac{\pi m_n}{cL}\right)^2.
\end{equation}
Then, the following statements hold.
\begin{enumerate}
\item
There is a unique real parameter $\rho$ and an even real
potential $V\in L^1(I)$, unique almost everywhere on $I$,
such that the symmetric Robin problem
\begin{equation}\label{hgrp}
-\psi''+V\psi=\Lambda\psi,\qquad
\psi'(a)=\rho\psi(a),\qquad
\psi'(b)=-\rho\psi(b)
\end{equation}
has ordered spectrum $(\Lambda_n)_{n\geq 0}$. Moreover, $V$ has a continuous representative, which is uniquely determined at every point of $I$.
\item
The zero-energy ground state $s$ can be chosen even
and strictly positive with its normalization fixed by $s(0)=1$.
In the optical coordinate with constant velocity $v_0$,
every choice $\kappa>0$ gives a Luttinger profile $K(y)=\kappa s^2(y)$. The resulting massless TLL is reflection symmetric,
its original mode functions satisfy Neumann boundary
conditions and its oscillator energies are
\begin{equation}\label{hald_general_energies}
E_n=v_0\sqrt{\Lambda_n}
=\frac{\pi v_0}{cL}m_n.
\end{equation}
It exhibits oscillator PWT at $T=cL/v_0$ with shortest
oscillator-PWT time $T_{\min}=cL/(gv_0)$ where $g=\gcd\{m_n:n\geq 1\}$.
\item
Conversely, every regular reflection-symmetric massless
TLL with bounded Liouville potential and oscillator PWT
has a representative of this form in the optical coordinate.
For fixed $L$ and $v_0$, the remaining freedom consists of
the scale $\kappa$ and a regular increasing odd
reparametrization $y=y(x)$ mapping $I$ onto itself.
In physical coordinates, we have $K(x)=\kappa s^2(y(x))$ and $
v(x)=v_0/y'(x)$. These profiles have the prescribed optical crossing time $L/v_0$.
\end{enumerate}
\end{theorem}
\begin{proof}
Let us apply the form of Hald's inverse spectral theorem stated
in Appendix~\ref{app_hald}, using the free Neumann pair
$(\widetilde{V},\widetilde{\rho})=(0,0)$ as the comparison
problem. The target eigenvalues are strictly increasing
and equal $(\pi n/L)^2$ for all $n\geq N_\ast$. Hence, the theorem gives the unique symmetric pair $(V,\rho)$ and a continuous representative of $V$. Since $\Lambda_0=0$ is the lowest eigenvalue, its eigenfunction can be chosen strictly positive. Reflection symmetry and simplicity of this eigenvalue
then make the normalized ground state even. Furthermore, it satisfies $-s''+Vs=0$ with $s'(a)=\rho s(a)$ and $s'(b)=-\rho s(b)$. Hence, $V=s''/s$. Continuity of $V$ gives
$s\in C^2(I)$, so $K=\kappa s^2$ is a regular positive
profile. The substitution $\psi=su$ transforms $-(v_0^2/K)(Ku')'=E^2u$ with $u'(a)=u'(b)=0$ into \eqref{hgrp} with
$\Lambda=E^2/v_0^2$. Equation~\eqref{hald_general_energies}
and the parity of $m_n$ establish oscillator PWT at
$T=cL/v_0$. Furthermore, Proposition~\ref{minimality_gcd} gives
the shortest transfer time. Conversely, Theorem~\ref{ttq} places the spectrum of every channel in the stated class in the form \eqref{htsg}. Uniqueness identifies its Liouville potential and induced Robin parameter with the reconstructed pair. Its positive zero mode determines $K$ up to an overall
scale. Returning to physical coordinates through $y=y(x)$ gives the stated velocity and Luttinger profiles while preserving the spectrum.
\end{proof}
Uniqueness also extends the rigidity result at the
optical crossing time to the full symmetric Robin
class. Notice that transfer at this time fixes the entire spectrum to the free Neumann spectrum, which in turn determines
both the potential and the endpoint parameter.
\begin{corollary}
A regular reflection-symmetric massless TLL with bounded
Liouville potential and oscillator PWT at $T=L/v_0$ satisfies
$V=0$, $\rho=0$ and $K=\mathrm{const}$.
\end{corollary}
\begin{proof}
Corollary~\ref{gtfl} gives the complete free Neumann
spectrum. Uniqueness in Theorem~\ref{chr} then identifies
the reconstructed pair as $(V,\rho)=(0,0)$.
The positive zero mode satisfies $s''=0$ and
$s'(a)=s'(b)=0$, so $s$ and $K$ are constant.
\end{proof}
For the profiles reconstructed above, the first heat
coefficient relates the induced Robin parameter to
the spatial variation of $K$. As the following proposition
shows, this relation forces $\rho<0$ whenever $K$ is
nonconstant and bounds the ratio $K_{\max}/K_{\min}$
in terms of $L$ and $\rho$.
\begin{proposition}
Let $s\in C^2(I)$ be even and strictly positive.
For $\kappa>0$, we define $K=\kappa s^2$, $V=s''/s$ and $\rho=s'(a)/s(a)=-s'(b)/s(b)$. we further suppose that the resulting symmetric Robin spectrum agrees with the free Neumann spectrum for all sufficiently large mode indices. Then, we have
\begin{equation}\label{ebi}
\rho=-\frac{1}{2}\int_I\left(\frac{s'}s\right)^2~dy
=-\frac{1}{8}\int_I[(\log K)']^2~dy\leq 0
\end{equation}
with $\rho=0$ if and only if $K$ is constant. Let us Write $K_{\max}=\max_I K$ and $K_{\min}=\min_I K$. Then, we also have
\begin{equation}\label{ebc}
\log^2\left(\frac{K_{\max}}{K_{\min}}\right)\leq-2L\rho.
\end{equation}
\end{proposition}
\begin{proof}
For the symmetric Robin boundary conditions, the heat-trace expansion begins with
\begin{equation}\label{HTE}
\Theta_{V,\rho}(t)=\frac{L}{\sqrt{4\pi t}}+\frac{1}{2}
-\frac{t^{1/2}}{\sqrt{4\pi}}\left[\int_I V(y)~dy+4\rho\right]+O(t).
\end{equation}
Since the spectrum differs from the free Neumann spectrum
at only finitely many levels, the difference of the heat
traces is analytic at $t=0$. Hence, its $t^{1/2}$ coefficient
must therefore vanish, giving $\int_I V(y)~dy+4\rho=0$. Let us set $\sigma=s'/s$. Then, $V=\sigma'+\sigma^2$ and
$\sigma(a)=\rho$, $\sigma(b)=-\rho$, so $\int_I V(y)~dy
=-2\rho+\int_I\sigma(y)^2~dy$. Combining these identities gives \eqref{ebi}. In particular, $\rho=0$ precisely when $s$, and hence $K$, is constant. To estimate the ratio $K_{\max}/K_{\min}$, we use the evenness of $K$ to choose points $y_{\max},y_{\min}\in[0,L/2]$ at which its maximum
and minimum are attained. Finally, Cauchy--Schwarz yields
\begin{equation}
\log^2{\left(\frac{K_{\max}}{K_{\min}}\right)}
\leq\frac{L}{2}\int_0^{L/2}[(\log K)']^2~dy
=\frac{L}{4}\int_I[(\log K)']^2~dy
=-2L\rho
\end{equation}
and this proves \eqref{ebc}.
\end{proof}

\subsection{Explicit one-defect profiles and transfer times}

\begin{figure}[t]
\centering
\includegraphics[width=\linewidth]{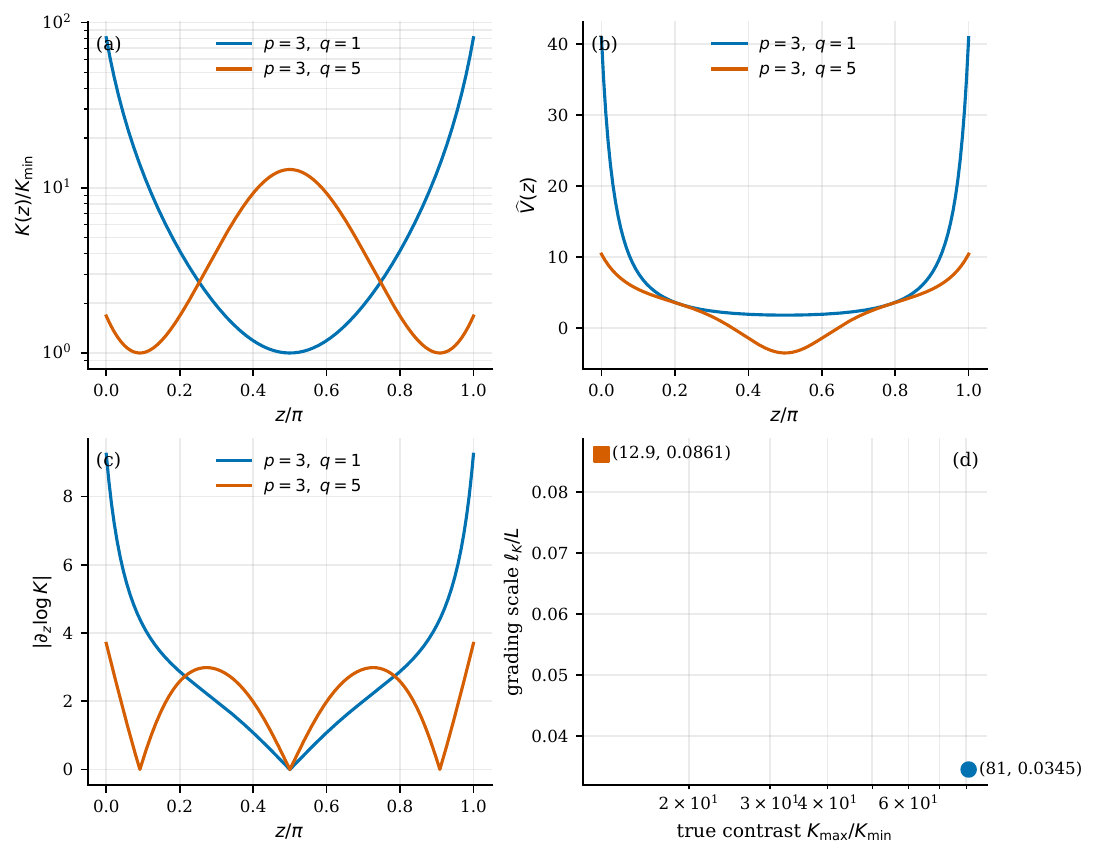}
\caption{Comparison of the profiles $r=3$ and $r=3/5$,
both with shortest oscillator-PWT time $3L/v_0$.
The panels show (a) normalized Luttinger profiles,
(b) dimensionless Liouville potentials,
(c) logarithmic gradients and (d) the ratios
$K_{\max}/K_{\min}$ and minimum variation lengths
$\ell_K=[\max_y|\partial_y\log K|]^{-1}$.
The ratio $K_{\max}/K_{\min}$ is $81$ for $r=3$
and approximately $12.913$ for $r=3/5$. The curves are obtained
from the explicit formulas in Theorem~\ref{rodf}.}
\label{Rdc}
\end{figure}

\begin{figure}[t]
\centering
\includegraphics[width=\linewidth]{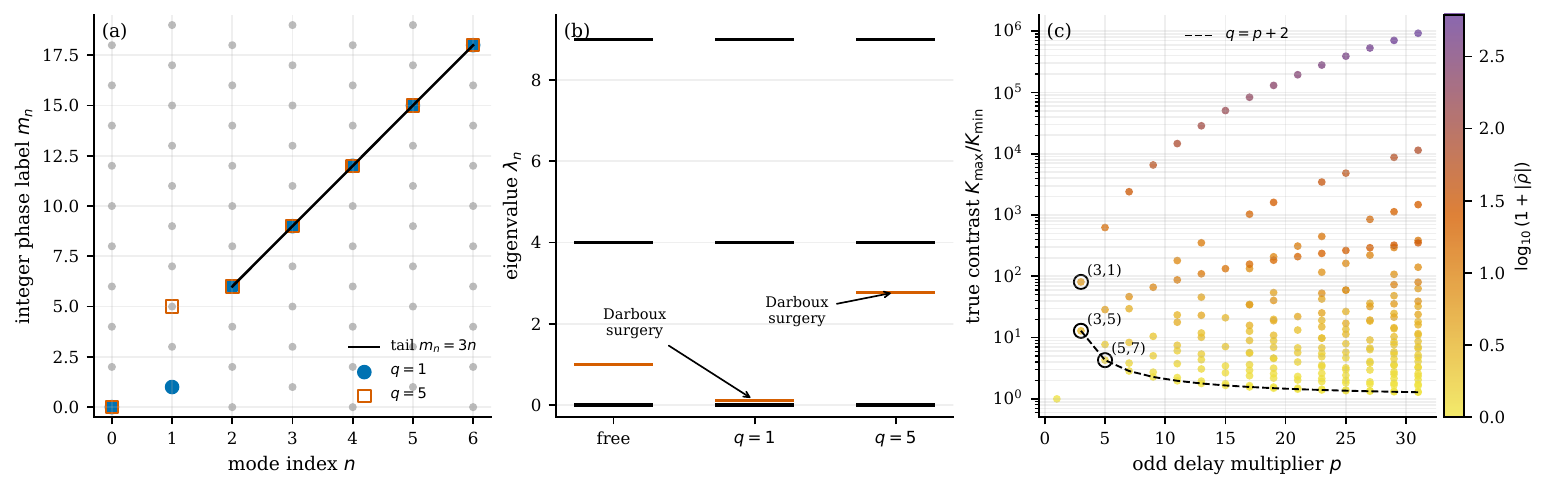}
\caption{Arithmetic conditions and profile variation
in the one-defect family. (a) The choices $q=1$ and $q=5$ share the phase-integer tail $m_n=3n$ and the shortest oscillator-PWT time $3L/v_0$. (b) The Darboux construction replaces the first positive eigenvalue while leaving all higher
eigenvalues unchanged. (c) Admissible parameters $r=p/q$ have coprime positive odd integers $p,q$ with $q<2p$. Color denotes the numerically evaluated ratio $K_{\max}/K_{\min}$.
Along the dashed sequence $q=p+2$, the profiles approach a constant as $p$ increases, while $T_{\min}=pL/v_0$ diverges.}
\label{adl}
\end{figure}

The general realization theorem presented in the previous Subsection becomes particularly explicit when only the first positive eigenvalue differs from the free Neumann spectrum. We now construct the complete regular family with this spectral form and determine which members exhibit oscillator PWT. The parameter $r>1/2$ controls the shifted eigenvalue, while the transfer condition selects rational values with odd numerator and denominator. The odd-integer family is a restricted subfamily of this construction.
\begin{theorem}\label{rodf}
Let $r>1/2$ and define $z=\pi (y-a)/L\in[0,\pi]$, $\vartheta_r(z)=(z-\pi/2)/r$ and $a_r=\pi/(2r)$. Furthermore, we set
\begin{equation}
D_r(z)=r\sin{z}\cos{\vartheta_r(z)}-\cos{z}\sin{\vartheta_r(z)},\qquad
N_r(z)=r^{-1}\sin{z}\cos{\vartheta_r(z)}-\cos{z}\sin{\vartheta_r(z)}.
\end{equation}
Both functions are strictly positive on $[0,\pi]$ and
symmetric about $z=\pi/2$. With primes denoting derivatives
with respect to $z$, we define
\begin{eqnarray}
\widehat{V}_r&=&-2(\log{D_r})'',\qquad
s_r=\frac{N_r}{D_r},\qquad
K_r(y)=\kappa s_r^2(z(y)),\label{rational_family_fields}\\
\widehat{\rho}_r&=&\frac{(1-r^2)\cos{a_r}}{r\sin{a_r}},\qquad
\rho_r=\frac{\pi}{L}\widehat{\rho}_r,\qquad
V_r(y)=\left(\frac{\pi}{L}\right)^2\widehat{V}_r(z(y)),\label{rational_family_endpoint_data}
\end{eqnarray}
where $\kappa>0$ is arbitrary. These functions are smooth and  $K_r$ is strictly positive. Moreover, $\widehat{V}_r=s_r''/s_r$. At $r=1$, they give $D_1=N_1=1$, $K_1=\kappa$ and $V_1=\rho_1=0$. The symmetric Robin problem \eqref{hgrp} with $(V,\rho)=(V_r,\rho_r)$ has ordered spectrum
\begin{equation}\label{rational_one_defect_spectrum}
\Lambda_0=0,\qquad
\Lambda_1=\left(\frac{\pi}{rL}\right)^2,\qquad
\Lambda_n=\left(\frac{\pi n}{L}\right)^2
\end{equation}
for $n\geq 2$.
\end{theorem}
\begin{proof}
Reflection $z\mapsto\pi-z$ sends $\vartheta_r\mapsto-\vartheta_r$ and leaves $D_r$ and $N_r$ unchanged. It therefore suffices to establish positivity on $[0,\pi/2]$. Direct differentiation gives
\begin{equation}
D_r'=\frac{r^2-1}{r}\cos{z}\cos{\vartheta_r},\qquad
N_r'=\frac{r^2-1}{r^2}\sin{z}\sin{\vartheta_r}
\end{equation}
and the endpoint values are
\begin{equation}
D_r(0)=N_r(0)=\sin{\left(\frac{\pi}{2r}\right)}>0,\qquad
D_r(\pi/2)=r,\qquad
N_r(\pi/2)=\frac{1}{r}.
\end{equation}
For $r>1$, $D_r$ increases and $N_r$ decreases between
positive endpoint values. For $1/2<r<1$, $N_r$ increases,
while $D_r'$ changes sign once from positive to negative.
Thus, $D_r$ has an interior maximum and its minimum
is attained at an endpoint. At $r=1$, both functions
are identically one. This proves their positivity
and the regularity of the profiles. Let us introduce the dimensionless spectral parameter $\mu=(L/\pi)^2\Lambda$. The endpoint calculation in Appendix~\ref{app_cc}, with its parameter $c$ replaced by $r$, gives the characteristic function
\begin{equation}
\Omega_r(\mu)=\frac{\mu-r^{-2}}{\mu-1}
\sqrt{\mu}\sin{(\pi\sqrt{\mu})}.
\end{equation}
For $r=1$, this reduces to the free Neumann
characteristic function. For $r\neq 1$, the apparent
singularity at $\mu=1$ is removable with $\Omega_r(1)=-\pi(1-r^{-2})/2$. We immediately see that its zeros are $0,r^{-2},4,9,\ldots$. Moreover, the condition $r>1/2$ places the shifted eigenvalue strictly between zero and four, giving
\eqref{rational_one_defect_spectrum}. The transformed zero-energy solution is $N_r/D_r$, which establishes $\widehat{V}_r=s_r''/s_r$. Uniqueness of the symmetric Robin inverse problem then identifies these profiles as the complete family with the stated spectrum, up to the overall scale
of $K$ and the coordinate freedom described in Theorem~\ref{chr}.
\end{proof}
Notice that the construction is regular for every $r>1/2$. Oscillator PWT imposes the additional requirement
that the shifted first mode acquire its reflection
phase at the same time as the unchanged higher modes.
This constraint gives the following arithmetic
condition on $r$.
\begin{corollary}\label{oorf}
A member of Theorem~\ref{rodf} exhibits oscillator PWT if and only if $r=p/q$ in lowest terms, where $p$ and $q$ are positive odd integers satisfying $q<2p$. Its shortest oscillator-PWT time and the corresponding phase integers are
$T_{\min}=pL/v_0$, $m_1=q$, $m_n=pn$ for $n\geq 2$.
\end{corollary}
\begin{proof}
The unchanged higher modes require $c=Tv_0/L$ to be a positive odd integer. The first mode requires $c/r$ to be a positive
odd integer as well. Hence, $r$ must be a ratio of odd integers. Let us write $r=p/q$ in lowest terms. Since $cq/p$ is an integer and $p,q$ are coprime, $c$ is an integer multiple of $p$. If we write $c=pk$, the phase conditions require both $pk$ and $kq$ to be odd, which imply that $p,q,k$ are odd. The smallest positive choice is $k=1$. The resulting sequence $\{q,2p,3p,\ldots\}$ has greatest common divisor one. This confirms that $T_{\min}=pL/v_0$.
\end{proof}
Different denominators can therefore produce the same
shortest transfer time while changing the spatial
variation of $K$. The choice $r=3/5$ provides a useful
comparison with the integer value $r=3$.
\begin{example}\label{ex1}
For $(p,q)=(3,5)$, the first positive frequency is $5\pi v_0/(3L)$ and the shortest oscillator-PWT time is $3L/v_0$. At this time, the phase integers are $m_1=5$ and $m_n=3n$ for $n\geq 2$, so every mode acquires the phase required by its
reflection parity. The endpoint parameter and the endpoint-to-midpoint ratio are $\widehat{\rho}_{3/5}=-16\sqrt{3}/15$ and $
K_{3/5}(a)/K_{3/5}((a+b)/2)=81/625$, respectively. Hence, the midpoint value is approximately $7.71605$ times the endpoint value. The minimum of $K_{3/5}$ lies in the interior, slightly away from the endpoint, so the ratio over the full interval is larger. Numerical evaluation of the explicit profile gives $K_{\max}/K_{\min}\approx12.9129528750$. For $r=3$, the corresponding ratio is $81$. Thus, the two profiles have the same shortest transfer time and substantially different ranges
of Luttinger-parameter values.
\end{example}
Figure~\ref{Rdc} compares these profiles, their
Liouville potentials and their spatial gradients. The integer family follows by setting $r=c$ with odd $c\geq 3$, giving shortest transfer time $cL/v_0$ and ratio $K_{\max}/K_{\min}=c^4$. Appendix~\ref{app_rod} gives the Darboux--Crum derivation, mode functions and endpoint formulas for the restricted range $r>1$.
The rational family also contains profiles arbitrarily
close to a constant Luttinger parameter. The sequence $r_p=p/(p+2)$ makes this limit explicit and relates the decreasing spatial variation to an increasing shortest transfer time.
\begin{corollary}
For every odd integer $p\geq 3$, we set $q=p+2$ and $r_p=p/(p+2)$. Then, $T_{\min}=pL/v_0$ and as $p\to\infty$,
\begin{equation}
\frac{K_{r_p}((a+b)/2)}{K_{r_p}(a)}
=\left(1+\frac{2}{p}\right)^4=1+\frac{8}{p}+\frac{24}{p^2}+O\left(\frac{1}{p^3}\right),\qquad
\widehat{\rho}_{r_p}=-\frac{4\pi}{p^2}+\frac{4\pi}{p^3}+O\left(\frac{1}{p^4}\right).
\end{equation}
If the profiles are normalized to a common midpoint value
$K_\ast>0$, then, as $p\to\infty$ through odd integers,
$K_{r_p}\to K_\ast$ and $V_{r_p}\to 0$ in $C^\infty(I)$. Thus, nonconstant profiles with exact oscillator PWT approach the constant profile smoothly, while their shortest transfer times diverge.
\end{corollary}
\begin{proof}
The integers $p$ and $p+2$ are coprime because $p$ is odd and $p+2<2p$ for $p\geq 3$. Hence, Corollary~\ref{oorf} gives the stated transfer time. The expansions follow from \eqref{rational_family_fields} and \eqref{rational_family_endpoint_data}. Since the explicit formulas depend analytically on $r$ near $r=1$ and their denominators remain uniformly bounded away from zero on $I$, the convergence is uniform together with every spatial derivative.
\end{proof}
Because the construction changes a single eigenvalue,
its heat trace differs from the free trace by just one pair of exponential terms. This gives an exact relation between the spectral deformation and the local heat coefficients discussed above.
\begin{proposition}
Let $\Theta_r(t)$ be the heat trace of the Robin operator in Theorem~\ref{rodf}. Moreover, let $\Theta_0^N(t)$ be the free Neumann heat trace. For every $t>0$, we have
\begin{equation}\label{erht}
\Theta_r(t)-\Theta_0^N(t)=e^{-t(\pi/L)^2r^{-2}}-e^{-t(\pi/L)^2}.
\end{equation}
In particular,
\begin{equation}\label{rlhi}
\rho_r^2-\frac{V_r(a)+V_r(b)}4=\left(\frac{\pi}{L}\right)^2\left(1-\frac{1}{r^2}\right).
\end{equation}
\end{proposition}
\begin{proof}
The two spectra differ only by the replacement of
$(\pi/L)^2$ with $(\pi/L)^2r^{-2}$, leading to \eqref{erht}.
The coefficient of $t$ on its right-hand side is
$(\pi/L)^2(1-r^{-2})$. If we equate it with the coefficient in the symmetric Robin heat expansion, we obtain \eqref{rlhi}.
\end{proof}
The same comparison applies at higher orders. Since the right-hand side of \eqref{erht} contains only integer powers of $t$, the coefficient of $t^{3/2}$ gives
\begin{equation}
\frac{1}{2}\int_a^b V_r(y)^2~dy+\frac{1}{2}\left[V_r'(b)-V_r'(a)\right]+2\rho_r\left[V_r(a)+V_r(b)\right]-\frac{8}{3}\rho_r^3=0.
\end{equation}
Here, primes denote derivatives with respect to $y$. For $\rho_r=0$, the expression reduces to the Neumann coefficient in \eqref{nhte}. Figure~\ref{adl} brings together the phase condition, the single-level
spectral replacement and the admissible rational
parameters. It also shows how the ratio
$K_{\max}/K_{\min}$ varies across the family and
along the sequence approaching the constant profile.
The transfer conditions established so far concern the oscillator modes. The overall scale of $K$ provides an additional parameter with which to align the compact zero-mode phases. Since this scale leaves the oscillator spectrum unchanged, it can be chosen to produce global PWT at any
selected oscillator-PWT time.
\begin{proposition}
Let us suppose positive profiles $v$ and $K_0$ define an
oscillator-PWT channel at time $T$. Moreover, we set
\begin{equation}
W_0=\int_I\frac{K_0(x)}{v(x)}~dx.
\end{equation}
For $K_\kappa=\kappa K_0$ with $\kappa>0$, the mode equation, Liouville potential, Robin parameter and oscillator spectrum are unchanged, while $W_\kappa=\kappa W_0$. For the integer charge lattice and canonical scalar lift, choosing
\begin{equation}\label{gps}
\kappa=\frac{T}{4\ell W_0}
\end{equation}
with $\ell\in\mathbb N$ gives global PWT at the same time $T$.
\end{proposition}
\begin{proof}
Both coefficients $vK$ and $K/v$ of the weighted mode problem acquire the factor $\kappa$, which cancels from the differential expression. The physical Neumann boundary conditions are also unchanged. Likewise, multiplication of
$\sqrt{K_0}$ by $\sqrt{\kappa}$ leaves the ratios $s''/s$ and $s'/s$ unchanged. Finally, \eqref{gps} gives $T/(4W_\kappa)=\ell$, satisfying the zero-mode condition in Theorem~\ref{zmc}.
\end{proof}
The explicit profiles make the dependence of delayed transfer on the induced endpoint conditions accessible to quantitative comparison. Their spectra, Robin parameters and transfer times follow from the continuum construction, while the free scale of $K$ controls the additional zero-mode phase alignment.
These exact solutions provide the reference profiles
for the finite-site convergence and sensitivity
studies that follow.

\section{Finite-band robustness and dispersive corrections}
\label{finite_band_robustness}

The exact continuum transfer identities provide a reference
for assessing implementations within the finite range of
energies described by the TLL approximation. In this section,
we examine how transfer accuracy depends on the accessible
modes, perturbations of the velocity and Luttinger profiles,
and dispersive corrections beyond the linearized theory. For bosons, a uniform bound on the many-body transfer error
requires an occupation or energy constraint in addition
to a modal cutoff. Even within a finite set of modes,
unbounded occupation can accumulate a small single-mode
phase error into a finite transfer error. We therefore
begin with exact finite-band identities and establish
error bounds under explicit constraints on the states.
We then derive the first variation under perturbations
of $v$ and $K$ and analyze the contribution of ultraviolet
dispersion. The estimates based on modal phase errors assume reflection symmetry, so that the one-particle Hamiltonian commutes with parity. We address perturbations that break this
symmetry separately below.

\subsection{Transfer errors under mode and occupation constraints}

Let $h$ be the positive one-particle oscillator Hamiltonian
and $\mathsf p$ the one-particle reflection operator. Reflection symmetry allows us to choose the eigenmodes $u_n$
with definite parity, so that $hu_n=E_nu_n$ and $\mathsf p u_n=p_nu_n$, where $p_n=(-1)^n$ for $n\geqslant1$. We remove the state-independent vacuum phase in such a way that 
both $U_{\mathrm{osc}}(T)$ and $P_{\mathrm{osc}}$ fix the oscillator vacuum. Let us define the projection onto the first $N$ nonzero modes by
\begin{equation}
\mathsf{P}_N:=\sum_{n=1}^N|u_n\rangle\langle u_n|.
\end{equation}
For each time $T$, let $\delta_n(T)\in(-\pi,\pi]$ measure the deviation of the dynamical phase from the reflection phase
\begin{equation}
e^{-iE_nT}=p_n e^{-i\delta_n(T)}.
\end{equation}
We may therefore write $E_nT=\pi j_n+\delta_n(T)$, where
$j_n$ is the unique integer of the same parity as $n$ for which $\delta_n(T)\in(-\pi,\pi]$. Let us set
\begin{equation}
\Delta_N(T):=\max_{1\leq n\leq N}|\delta_n(T)|.
\end{equation}
On the one-particle band, the largest phase deviation
determines the transfer error exactly and provides
a bound uniform over all normalized states in that band.
\begin{proposition}\label{opfbe}
The one-particle transfer error on the first $N$ modes
satisfies
\begin{equation}\label{errineq}
\varepsilon_N^{(1)}(T):=\left\|
\left(e^{-ihT}-\mathsf{p}\right)\mathsf{P}_N
\right\|=2\max_{1\leq n\leq N}
\left|\sin{\frac{\delta_n(T)}2}\right|
\leq\Delta_N(T).
\end{equation}
\end{proposition}
\begin{proof}
The operator $(e^{-ihT}-\mathsf{p})\mathsf{P}_N$ vanishes on the orthogonal complement of $\operatorname{Ran}\mathsf{P}_N$. On $\operatorname{Ran}\mathsf{P}_N$, it is diagonal in the orthonormal
basis $\{u_n\}_{n=1}^N$, with eigenvalues
\begin{equation}
e^{-iE_nT}-p_n=p_n\left(e^{-i\delta_n(T)}-1\right).
\end{equation}
Since $|p_n|=1$, their moduli are $\left|e^{-iE_nT}-p_n\right|
=2\left|\sin{\left(\delta_n(T)/2\right)}\right|$. The operator norm is the largest of these moduli, which gives equality in \eqref{errineq}. Finally, $2|\sin{(x/2)}|\leq |x|$ bounds this maximum by $\Delta_N(T)$.
\end{proof}
For a state with occupations $k_1,\ldots,k_N$, the total
phase deviation is $\sum_{n=1}^N k_n\delta_n(T)$, so
the transfer error depends on both the modal phase
deviations and the occupation numbers. Let $\mathcal{F}_N$ be the bosonic Fock space over $\operatorname{Ran}\mathsf{P}_N$. Moreover, we define $N_n=a_n^\dagger a_n$ and $\mathcal{N}_N=\sum_{n=1}^N N_n$. For a positive integer $M$, let $\mathsf{P}_{N,M}$ project onto the subspace with total occupation at most $M$, that is
\begin{equation}
\mathcal{F}_{N,M}:=\operatorname{span}
\left\{|k_1,\ldots,k_N\rangle
\ \middle|\
k_n\in\mathbb{N}_0,~\sum_{n=1}^N k_n\leq M\right\}.
\end{equation}
The occupation constraint controls the accumulated
phase error and gives the following extension of
the one-particle estimate.
\begin{theorem}\label{tfbne}
On $\mathcal{F}_{N,M}$, the oscillator transfer error is
\begin{equation}\label{efce}
\varepsilon_{N,M}^{\mathrm{F}}(T):=
\left\|\left(U_{\mathrm{osc}}(T)-P_{\mathrm{osc}}\right)
\mathsf{P}_{N,M}\right\|
=\max_{\substack{k_1,\ldots,k_N\in\mathbb{N}_0\\
  \sum_n k_n\leq M}}\left|
\exp\left(-i\sum_{n=1}^N k_n\delta_n(T)\right)-1\right|.
\end{equation}
The transfer error is therefore bounded below by the
one-particle error and above by the maximal accumulated
phase deviation. More precisely, we have
\begin{equation}\label{efeh}
\varepsilon_N^{(1)}(T)\leq\varepsilon_{N,M}^{\mathrm{F}}(T)
\leq\min\{2,M\Delta_N(T)\}.
\end{equation}
\end{theorem}
\begin{proof}
On the occupation vector $|k_1,\ldots,k_N\rangle$, the dynamical phase divided by the reflection phase is
\begin{equation}
\frac{\exp(-iT\sum_n k_n E_n)}{\prod_n p_n^{k_n}}=
\exp\left(-i\sum_n k_n \delta_n(T)\right).
\end{equation}
Both unitaries are diagonal in this basis, so their
difference has the norm in \eqref{efce}. Moreover, the upper bound follows from
\begin{equation}
|e^{-ix}-1|\leq\min\{2,|x|\},\qquad
\left|\sum_n k_n \delta_n(T)\right|\leq M\Delta_N(T),
\end{equation}
while the one-particle subspace is contained in $\mathcal {F}_{N,M}$, thus giving the lower bound.
\end{proof}
The dependence on $M$ reflects the accumulation of phase errors with occupation. To see why an occupation constraint matters, we consider a mode with nonzero defect $\delta$ and allow its occupation to grow without bound. Then, we obtain
\begin{equation}
\left\|\left.\left(U_{\mathrm{osc}}(T)-P_{\mathrm{osc}}\right)
\right|_{\mathcal{F}_N}\right\|
\geq\sup_{k\geq 0}|e^{-ik\delta}-1|.
\end{equation}
If $\delta/(2\pi)$ is irrational, the phases $e^{-ik\delta}$ are dense on the unit circle and the supremum is two. For $\delta/(2\pi)=p/q$ in lowest terms, these phases run through the $q$-th roots of unity. The supremum is two when $q$ is even and $2\cos{(\pi/(2q))}$ when $q$ is odd. Thus, arbitrarily small nonzero phase defects can produce a finite, unrestricted Fock-space transfer error. By contrast, for fixed $M$, \eqref{efeh} ensures that the error tends to zero with $\Delta_N(T)$. An alternative to a strict occupation cutoff is
to control the occupation moments of the state. On $\mathcal{F}_N$, we define $D_N(T):=\sum_{n=1}^N\delta_n(T) N_n$. Then, $U_{\mathrm{osc}}(T)=P_{\mathrm{osc}}e^{-iD_N(T)}$.
For a normalized state $|\Psi\rangle$ with $\langle\Psi|\mathcal{N}_N^2|\Psi\rangle<\infty$, the spectral inequality $|e^{-ix}-1|\leq|x|$ gives
\begin{equation}\label{ncsb}
\left\|\left(U_{\mathrm{osc}}(T)-P_{\mathrm{osc}}\right)|\Psi\rangle\right\|\leq\|D_N(T)|\Psi\rangle\|\leq\Delta_N(T)\sqrt{\langle\Psi|\mathcal{N}_N^2|\Psi\rangle}.
\end{equation}
This estimate controls the error through the second
moment of the total occupation. The state-dependent error bound also gives a lower bound on the transfer amplitude. To see that, we denote the right-hand side of \eqref{ncsb} by $\eta_\Psi$. Since the evolved and reflected states are normalized, we have
\begin{equation}
\operatorname{Re}\langle\Psi|P_{\mathrm{osc}}^\dagger U_{\mathrm{osc}}(T)|\Psi\rangle=1-\frac{1}{2}\left\|
\left(U_{\mathrm{osc}}(T)-P_{\mathrm{osc}}\right)|\Psi\rangle
\right\|^2.
\end{equation}
Combining this identity with \eqref{ncsb} yields
\begin{equation}
\left|\langle\Psi|P_{\mathrm{osc}}^\dagger U_{\mathrm{osc}}(T)
|\Psi\rangle\right|\geq\max\left\{0,1-\frac{\eta_\Psi^2}{2}\right\}.
\end{equation}

\subsection{Charge-constrained robustness of global PWT}

The preceding estimates concern oscillator transfer. For global PWT, the compact zero-mode phases must also align across the charge sectors used to encode the state. We continue to assume the integer charge lattice, the canonical scalar lift of reflection, and removal of the state-independent vacuum phase. Let $T_0$ be an exact global-PWT time for a reference profile with integrated weight $W_0$. Then, we have $\ell:=T_0/(4W_0)\in\mathbb{N}$. For perturbed values $T$ and $W$, we define the zero-mode phase deviation in charge sector $q\in\mathbb{Z}$ by
\begin{equation}
\zeta_q(T,W):=\frac{\pi q^2}{2}\left(\frac{T}{W}-\frac{T_0}{W_0}\right).
\end{equation}
Here $\zeta_q(T,W)$ is evaluated without reduction modulo
$2\pi$, so that its dependence on $T$ and $W$ remains explicit. Let $\Pi_Q^{\mathrm{zm}}$ project onto the charge
states with $|q|\leq Q$, where $Q\in\mathbb{N}_0$. Together with the mode and occupation constraints introduced above, this
charge cutoff gives a uniform bound on the combined transfer error. Moreover, we also assume that the perturbations preserve reflection symmetry.
\begin{proposition}
Let us define the global transfer error on $\operatorname{Ran}\Pi_Q^{\mathrm{zm}}\otimes\mathcal F_{N,M}$ by
\begin{equation}
\varepsilon_{Q;N,M}^{\mathrm{glob}}(T):=\left\|\left(U(T)-P\right)\left(\Pi_Q^{\mathrm{zm}}\otimes\mathsf{P}_{N,M}\right)
\right\|.
\end{equation}
Then, we have
\begin{equation}\label{ecge}
\varepsilon_{Q;N,M}^{\mathrm{glob}}(T)=\max_{\substack{q\in\mathbb{Z},~|q|\leq Q\\
k_1,\ldots,k_N\in\mathbb{N}_0,~\sum_n k_n\leq M}}\left|
e^{-i[\zeta_q(T,W)+\sum_{n=1}^N k_n\delta_n(T)]}-1
\right|.
\end{equation}
The charge and occupation constraints therefore give
\begin{equation}\label{cgeb}
\varepsilon_{Q;N,M}^{\mathrm{glob}}(T)\leq\min\left\{
2,~\frac{\pi Q^2}{2}\left|\frac{T}{W}-\frac{T_0}{W_0}\right|
+M\Delta_N(T)\right\}.
\end{equation}
\end{proposition}
\begin{proof}
At the reference time, we find
\begin{equation}
\exp\left(-\frac{i\pi T_0 q^2}{2W_0}\right)=e^{-2\pi i\ell q^2}=1.
\end{equation}
Hence, at the perturbed time, a basis state with charge $q$ and occupations $k_1,\ldots,k_N$ acquires, relative to reflection, the phase $\exp(-i\zeta_q(T,W)-i\sum_{n=1}^N k_n\delta_n(T))$. Notice that the evolution and reflection operators are diagonal in this basis, so the norm of their difference is the maximum in \eqref{ecge}. Using the inequalities
\begin{equation}
|\zeta_q(T,W)|\leq\frac{\pi Q^2}{2}\left|\frac{T}{W}-\frac{T_0}{W_0}\right|,\qquad
\left|\sum_n k_n\delta_n(T)\right|\leq M\Delta_N(T)
\end{equation}
together with $|e^{-ix}-1|\leq\min\{2,|x|\}$ gives \eqref{cgeb}.
\end{proof}
Within a fixed charge sector, $\zeta_q$ contributes a common
phase to all states and therefore has no effect on the
evolution of observables acting within that subspace. Across different charge sectors, its charge dependence changes the relative phases of coherent superpositions. The factor $Q^2$ in \eqref{cgeb} quantifies the resulting sensitivity to errors in $T/W$. For the smooth profile perturbations introduced in \eqref{pp} below, let $\alpha$ and $\beta$ denote the relative first-order changes in $v$ and $K$, respectively. Since $w=K/v$, the integrated weight satisfies
\begin{equation}
\dot{W}:=\left.\frac{dW_\eta}{d\eta}\right|_{\eta=0}=
\int_I w(x)\left[\beta(x)-\alpha(x)\right]~dx,
\end{equation}
where $w$ is evaluated at the reference profile. If the transfer time is adjusted according to $T_\eta=T_0+\eta\dot{T}+O(\eta^2)$, the zero-mode
term in \eqref{cgeb} becomes $2\pi\ell Q^2|\eta|\left|
(\dot{T}/T_0-(\dot{W}/W_0)\right|+O(\eta^2)$. Hence, preserving zero-mode alignment to first order requires
$\dot{T}/T_0=\dot{W}/W_0$. If we adjust the transfer time to reduce oscillator phase errors, a residual zero-mode phase mismatch may affect the relative phases between charge sectors while leaving transfer within each sector unchanged.

\subsection{First-order response to profile perturbations}
\label{fixed_hilbert_perturbations}

We now express the transfer error in terms of perturbations of the velocity and Luttinger profiles. In this subsection, we strengthen the regularity assumptions in \eqref{sr} to $v,K\in W^{2,\infty}(I)$ and consider
\begin{equation}\label{pp}
v_\eta(x)=v(x)\left[1+\eta\alpha(x)\right],\qquad
K_\eta(x)=K(x)\left[1+\eta\beta(x)\right],
\end{equation}
where $\alpha,\beta\in W^{2,\infty}(I)$ are real. The reference profiles are strictly positive and $|\eta|$ is chosen small enough to preserve positivity. Moreover, dots denote derivatives with respect to $\eta$ evaluated at $\eta=0$. If we write $p=vK$ and $w=K/v$, we have $p_\eta=v_\eta K_\eta$, $w_\eta=K_\eta/v_\eta$, $\dot{p}=p(\alpha+\beta)$ and $\dot{w}=w(\beta-\alpha)$. The squared mode energies are the eigenvalues of the generalized Neumann problem
\begin{equation}
-\frac{d}{dx}\left(p_\eta\frac{du}{dx}\right)=\lambda w_\eta u,\qquad
\left.p_\eta u'\right|_{\partial I}=0.
\end{equation}
The natural Hilbert space $\Hil_\eta=L^2(I,w_\eta\,dx)$ depends on the perturbation. To compare the corresponding evolutions, we transport them to a fixed space through the unitary map $J_\eta:\Hil_\eta\longrightarrow\Hil_{\mathrm{fix}}:=L^2(I,dx)$ such that $J_\eta u=\sqrt{w_\eta}~u$. The transported operators are
\begin{equation}\label{fhslo}
B_\eta:=J_\eta A_\eta J_\eta^{-1}=w_\eta^{-1/2}\left(-\frac{d}{dx}p_\eta\frac{d}{dx}\right)w_\eta^{-1/2},\qquad
h_\eta:=B_\eta^{1/2}.
\end{equation}
Their quadratic forms are
\begin{equation}
\mathfrak{b}_\eta[f,g]=\int_I p_\eta\left(\frac{f}{\sqrt{w_\eta}}\right)'\overline{\left(\frac{g}{\sqrt{w_\eta}}\right)'}~dx
\end{equation}
with $f,g\in H^1(I)$. This formulation places the perturbation on the common form domain $H^1(I)$, while the transformed boundary conditions enter through the domains of the associated self-adjoint operators~\cite{Kato1995}. Moreover, the finite-band propagator estimates below use
Lemma~\ref{lfbpr}. Let $f_n=\sqrt{w} u_n$ be the real normalized eigenfunctions of $B=B_0$. On the fixed space,
we write $(\mathsf{p} f)(x)=f(-x)$. The reference profiles are reflection symmetric, so $\mathsf{p} f_n=p_nf_n$ with $p_n=(-1)^n$. If we differentiate the quadratic form , we obtain 
\begin{equation}
\mathsf{M}_{mn}:=\dot{\mathfrak{b}}_0[f_m,f_n]
=\int_I p(\alpha+\beta)u_m' u_n'~dx-\frac{\lambda_m+\lambda_n}{2}\int_I w(\beta-\alpha)u_mu_n~dx.
\end{equation}
In particular, the generalized Hellmann--Feynman formula yields
\begin{equation}\label{foHFes}
\dot{\lambda}_n=\mathsf{M}_{nn}=\int_I\left[p(\alpha+\beta)(u_n')^2-\lambda_n w(\beta-\alpha)u_n^2\right]~dx,\qquad
\dot{E}_n=\frac{\dot\lambda_n}{2E_n},
\end{equation}
where the energy formula applies for $n\geq 1$. These expressions hold for arbitrary real $\alpha$ and $\beta$ satisfying the stated regularity assumptions. Throughout the following analysis, $\mathsf{P}_N=\sum_{n=1}^N |f_n\rangle\langle f_n|$ projects onto the first $N$ positive reference modes. The input band is fixed as $\eta$ varies, while the output is measured in the full space $\Hil_{\mathrm{fix}}$.

\subsubsection{Perturbations preserving reflection symmetry}

We suppose first that $\alpha$ and $\beta$ are even. The perturbed profiles then remain reflection symmetric and the evolution commutes with $\mathsf{p}$. At an exact reference transfer time, this symmetry causes all off-diagonal first-order contributions to cancel. The remaining error is
determined by the energy shifts and the adjustment of the transfer time.
\begin{theorem}\label{thm_fotd}
Let us assume that the reference channel exhibits exact oscillator PWT at $T_0$, $e^{-iE_nT_0}=p_n=(-1)^n$ and $E_n=\sqrt{\lambda_n}$ for $n\geq1$. Moreover, let $\alpha, \beta$ in \eqref{pp} be even and $T_\eta=T_0+\eta\dot{T}$.
For every fixed reference band, we have
\begin{equation}\label{fode}
\left(e^{-ih_\eta T_\eta}-\mathsf{p}\right)\mathsf{P}_N
=\eta\mathsf{D}_{\mathrm{even}}(\dot{T})\mathsf{P}_N
+O_N(\eta^2)
\end{equation}
in operator norm, where, for $m\geq 0$ and $1\leq n\leq N$
\begin{equation}
\langle f_m,\mathsf{D}_{\mathrm{even}}(\dot{T})f_n\rangle
=-ip_n\left(T_0\dot{E}_n+\dot{T} E_n\right)\delta_{mn}.
\end{equation}
Hence, the error $\varepsilon_{N,\eta}^{(1)}(T):=\left\|
\left(e^{-ih_\eta T}-\mathsf{p}\right)\mathsf{P}_N\right\|$
satisfies
\begin{equation}\label{fopeg}
\varepsilon_{N,\eta}^{(1)}(T_\eta)=|\eta|\max_{1\leq n\leq N}
|T_0\dot{E}_n+\dot{T} E_n|+O_N(\eta^2).
\end{equation}
\end{theorem}
\begin{proof}
The first variation of $h=B^{1/2}$ has matrix elements $\langle f_m,\dot{h} f_n\rangle=\mathsf{M}_{mn}/(E_m+E_n)$, whenever $(m,n)\neq (0,0)$. The lowest eigenvalue remains exactly zero under the perturbation, so $\mathsf {M}_{00}=\dot{\lambda}_0=0$. Its inclusion in the output basis
therefore introduces no vanishing denominator when $n\geq 1$. On the reference band, Duhamel differentiation gives
\begin{equation}
\left.\frac{d}{d\eta}e^{-ih_\eta T_0}\right|_{\eta=0}
=-i\int_0^{T_0}e^{-ih(T_0-s)}\dot{h} e^{-ihs}~ds.
\end{equation}
Writing $U_\eta^{(1)}(T)=e^{-ih_\eta T}$ and $\dot{U}^{(1)}(T)=
\left.\partial_\eta U_\eta^{(1)}(T)\right|_{\eta=0}$, the off-diagonal matrix elements are
\begin{equation}\label{offddf}
\langle f_m,\dot{U}^{(1)}(T_0)f_n\rangle
=(p_n-p_m)\frac{\mathsf{M}_{mn}}{\lambda_n-\lambda_m},\qquad m\neq n,
\end{equation}
where $E_0=0$ and $p_0=1$. If the two modes have opposite parity, evenness of $\alpha$ and $\beta$ gives $\mathsf {M}_{mn}=0$. If instead they have the same parity, the factor $p_n-p_m$ vanishes. All off-diagonal first-order terms therefore vanish, leaving the diagonal entries $-iT_0 p_n\dot{E}_n$. Including the variation of the transfer time gives the combined entries $-ip_n(T_0\dot{E}_n+\dot{T} E_n)$, whose largest modulus determines the first-order transfer error. Finally, Lemma~\ref{lfbpr} supplies the operator-norm remainder in \eqref{fode}, completing the proof.
\end{proof}
The theorem also determines how much of the error can be removed by adjusting the transfer time. The minimum first-order error coefficient is
\begin{equation}
\mathcal{R}_N:=\inf_{\dot{T}\in\mathbb R}\max_{1\leq n\leq N}
|T_0\dot{E}_n+\dot{T} E_n|=\pi\inf_{\xi\in\mathbb{R}}
\max_{1\leq n\leq N} m_n\left|\frac{\dot{E}_n}{E_n}-\xi\right|,
\end{equation}
where $E_n T_0=\pi m_n$ and $\xi=-\dot{T}/T_0$.
A common relative energy shift can therefore be cancelled by a single timing adjustment. The residual error measures how far the relative shifts differ from a common value with weights
set by the phase integers. A bound requiring only the amplitudes of the profile perturbations follows from \eqref{foHFes}. More precisely, using
\begin{equation}
\int_I p(u_n')^2~dx=\lambda_n,\qquad
\int_I wu_n^2~dx=1,
\end{equation}
we obtain
\begin{equation}
\left|\frac{\dot{E}_n}{E_n}\right|\leq\frac{1}{2}\left(
\|\alpha+\beta\|_\infty+\|\beta-\alpha\|_\infty\right)
\leq\|\alpha\|_\infty+\|\beta\|_\infty.
\end{equation}
At the unchanged reference time, we find
\begin{equation}
\varepsilon_{N,\eta}^{(1)}(T_0)\leq\pi m_N|\eta|\left(\|\alpha\|_\infty+\|\beta\|_\infty\right)+O_N(\eta^2).
\end{equation}
At the first conformal transfer time, $m_N=N$, so the first-order bound grows linearly with the number of modes. Notice that this estimate uses only the uniform amplitudes of the perturbations. Moreover, their spatial structure enters through the integrals in \eqref{foHFes}.

\subsubsection{Perturbations breaking reflection symmetry}

For asymmetric perturbations, the change in the weight also affects how spatial reflection is represented on the fixed Hilbert space. Let $\mathcal{R}$ act on a mode profile by
$(\mathcal{R}u)(x)=u(-x)$. Transport through $J_\eta$ gives
\begin{equation}\label{tpr}
\mathcal{R}_\eta^{\mathrm{prof}}:=J_\eta\mathcal{R} J_\eta^{-1}
=M_{r_\eta}\mathsf{p},\qquad
r_\eta(x)=\sqrt{\frac{w_\eta(x)}{w_\eta(-x)}},
\end{equation}
where $M_{r_\eta}$ denotes multiplication by the function $r_\eta$ defined above. The fixed unitary reflection $\mathsf{p}$ corresponds instead to the weighted-space map
\begin{equation}
(J_\eta^{-1}\mathsf{p} J_\eta u)(x)=\sqrt{\frac{w_\eta(-x)}{w_\eta(x)}}u(-x).
\end{equation}
These two maps coincide precisely when $w_\eta$ is even. For a general positive weight, $\mathcal{R}_\eta^{\mathrm{prof}}$ is a bounded involution and it is unitary if and only if
$w_\eta$ is even almost everywhere. Hence, exact equality between the unitary evolution and this profile-reflection map on the full one-particle space requires an even weight. To compute the first variation of the transported reflection operator, we define $\gamma:=\beta-\alpha$ and $\gamma_{\mathrm{o}}(x):=[\gamma(x)-\gamma(-x)]/2$. Since the reference weight is even, \eqref{tpr} gives
\begin{equation}\label{prfv}
\mathcal{R}_\eta^{\mathrm{prof}}=\mathsf{p}+\eta M_{\gamma_{\mathrm{o}}}\mathsf{p}+O(\eta^2)
\end{equation}
in operator norm. We measure transfer relative to the reflected
profile through $\varepsilon_{N,\eta}^{\mathrm{prof}}(T):=
\left\|\left(e^{-ih_\eta T}-\mathcal{R}_\eta^{\mathrm{prof}}\right)\mathsf{P}_N\right\|$. 
Both the evolved and reflected profiles are thus compared after transport to the same Hilbert space. The following proposition expresses the first-order transfer error in terms of dynamical phase shifts, mode mixing, and the variation of the transported reflection operator.
\begin{proposition}
For arbitrary real $\alpha$ and $\beta$ satisfying the stated regularity assumptions, let $T_\eta=T_0+\eta\dot{T}$. Then, we have
\begin{equation}
\left(e^{-ih_\eta T_\eta}-\mathcal{R}_\eta^{\mathrm{prof}}\right)\mathsf{P}_N=
\eta\mathsf{F}(\dot{T})\mathsf{P}_N+O_N(\eta^2).
\end{equation}
For $m\geq 0$ and $1\leq n\leq N$, the first-order matrix elements are
\begin{equation}\label{sdm}
\langle f_m,\mathsf{F}(\dot{T})f_n\rangle=
\begin{cases}
-ip_n\left(T_0\dot{E}_n+\dot{T} E_n\right),& m=n,\\
\displaystyle
(p_n-p_m)\frac{\mathsf{M}_{mn}}{\lambda_n-\lambda_m}
-p_n\langle f_m,M_{\gamma_{\mathrm{o}}}f_n\rangle,& m\neq n.
\end{cases}
\end{equation}
The first variation of the transported reflection operator vanishes precisely when the odd part of gamma is zero, which means that the perturbed weight remains even to first order.
\end{proposition}
\begin{proof}
The Duhamel calculation leading to \eqref{offddf} applies to arbitrary $\alpha$ and $\beta$. Subtracting the first variation in \eqref{prfv} adds the matrix element $-\langle f_m,M_{\gamma_{\mathrm{o}}}\mathsf{p} f_n\rangle=-p_n\langle f_m,M_{\gamma_{\mathrm{o}}}f_n\rangle$. For $m=n$, this term vanishes because the odd part of gamma is integrated against the even function $|f_n|^2$. Finally, the operator-norm estimate for the higher-order terms follows from Lemma~\ref{lfbpr}.
\end{proof}
The failure of the dynamics to commute with the fixed spatial reflection can also be measured directly by the quantity $\chi_N(\eta):=\|[h_\eta,\mathsf{p}]\mathsf{P}_N\|$. Differentiating at the symmetric reference point gives $\chi_N(\eta)=|\eta|\|[\dot{h},\mathsf{p}]\mathsf{P}_N\|+o(|\eta|)$ with matrix elements
\begin{equation}
\langle f_m,[\dot{h},\mathsf{p}]f_n\rangle=(p_n-p_m)\frac{\mathsf{M}_{mn}}{E_m+E_n}.
\end{equation}
Notice that this diagnostic measures coupling between modes
of opposite parity. The additional term $M_{\gamma_{\mathrm {o}}}\mathsf{p}$ in \eqref{sdm} accounts for the first variation of the transported reflection operator. We conclude this part by observing that if $\alpha$ and $\beta$ are even, the opposite-parity couplings and the first variation of the transported reflection operator vanish, thus recovering Theorem~\ref{thm_fotd}. If they are purely odd, the diagonal integrals in \eqref{foHFes} are zero, so $\dot{E}_n=0$ for every positive mode. We observe that  the transfer error can nevertheless have a nonzero first-order contribution from mode mixing and the first variation of the transported reflection operator. Adjusting the transfer time changes only the diagonal term in \eqref{sdm}, so it cannot cancel a nonzero off-diagonal error. These contributions must therefore be included when assessing perturbations that break reflection
symmetry.

\subsection{Profile sensitivity of a conformal channel}

For a conformal reference channel, the first-order energy
shifts can be expressed directly in terms of the spatial
profiles of the perturbations. Let us suppose that $K(x)=K_0$ is constant while $v(x)$ is positive and even. Concerning the regularity conditions we refer the reader to the preceding subsection. In the reference optical coordinate defined in \eqref{lca}, the normalized nonzero modes are
\begin{equation}
u_n(x(y))=\sqrt{\frac{2v_0}{K_0 L}}\cos{\left(\frac{\pi n}{L}\left(y+\frac{L}{2}\right)\right)}
\end{equation}
with $n\geq 1$. Their energies and the first oscillator-PWT time are $E_n=\pi nv_0/L$ and $T_0=L/v_0$, respectively. Moreover, the normalization follows from $w(x)dx=(K_0/v_0)dy$. We express the perturbations in the same reference
coordinate, writing $\alpha(y)$ and $\beta(y)$ for
$\alpha(x(y))$ and $\beta(x(y))$. Furthermore, we define
\begin{equation}
\overline{\alpha}:=\frac{1}{L}\int_{-L/2}^{L/2}\alpha(y)~dy,\qquad
b_n:=\frac{1}{L}\int_{-L/2}^{L/2}\beta(y)\cos{\left(\frac{2\pi n}{L}\left(y+\frac{L}{2}\right)\right)}~dy.
\end{equation}
The mean velocity perturbation shifts all mode energies by the same relative amount, which can be compensated by adjusting the transfer time. The following corollary expresses the residual phase errors in terms of the coefficients $b_n$ associated with the Luttinger perturbation.
\begin{corollary}
Under the regularity assumptions of the preceding
subsection, we assume $\alpha$ and $\beta$ to be even. Under these hypotheses, the first-order relative shift of each positive mode energy is given by 
\begin{equation}\label{crshift}
\frac{\dot{E}_n}{E_n}=\overline{\alpha}-b_n
\end{equation}
for every $n\geq 1$. If we choose the transfer time to agree with the perturbed optical crossing time to first order, i.e.
$T_\eta=T_0\left(1-\eta\overline{\alpha}\right)+O(\eta^2)$, then for each fixed $N$ and every $1\leq n\leq N$, the phase defects computed from the perturbed energies satisfy
\begin{equation}
\delta_n(T_\eta)=-\pi n\eta b_n+O_N(\eta^2),\qquad
\varepsilon_{N,\eta}^{(1)}(T_\eta)=\pi|\eta|\max_{1\leq n\leq N}n|b_n|+O_N(\eta^2).
\end{equation}
Moreover, the Fourier coefficients and transfer error obey
\begin{equation}\label{W11Keb}
|b_n|\leq\frac{\|\beta'\|_{L^1(I)}}{2\pi n},\qquad
\varepsilon_{N,\eta}^{(1)}(T_\eta)\leq\frac{|\eta|}{2}\|\beta'\|_{L^1(I)}+O_N(\eta^2).
\end{equation}
If $\beta'(-L/2)=\beta'(L/2)=0$, then we have
\begin{equation}\label{W21Keb}
|b_n|\leq\frac{L}{4\pi^2n^2}\|\beta''\|_{L^1(I)},\qquad
|\delta_n(T_\eta)|\leq\frac{L|\eta|}{4\pi n}\|\beta''\|_{L^1(I)}+O_N(\eta^2).
\end{equation}
Here, primes denote derivatives with respect to $y$ and $\varepsilon_{N,\eta}^{(1)}$ is measured on the fixed reference band defined above.
\end{corollary}
\begin{proof}
If we substitute the normalized conformal modes into \eqref{foHFes}, we get
\begin{equation}
\frac{\dot{\lambda}_n}{\lambda_n}=\frac{2}{L}\int_I\alpha(y)~dy
-\frac{2}{L}\int_I\beta(y)\cos{\left(\frac{2\pi n}{L}\left(y+\frac{L}{2}\right)\right)}~dy.
\end{equation}
Since $\dot{E}_n/E_n=\dot{\lambda}_n/(2\lambda_n)$, this proves \eqref{crshift}. Moreover, the optical crossing time changes according to
\begin{equation}
\int_I\frac{dx}{v_\eta(x)}=T_0-\eta\int_I\frac{\alpha(x)}{v(x)}~dx+O(\eta^2)=T_0(1-\eta\overline{\alpha})+O(\eta^2).
\end{equation}
Replacing this adjusted transfer time and the first-order energy shifts gives, for each mode in the fixed band,
$E_n(\eta)T_\eta-\pi n=-\pi n\eta b_n+O_N(\eta^2)$. For a fixed band and sufficiently small $|\eta|$, these deviations lie in the principal phase interval. The transfer-error formula then follows from \eqref{fopeg}. Moreover, one integration by parts yields
\begin{equation}
b_n=-\frac{1}{2\pi n}\int_I\beta'(y)\sin{\left(\frac{2\pi n}{L}\left(y+\frac{L}{2}\right)\right)}~dy.
\end{equation}
The sine vanishes at both endpoints, thus giving \eqref{W11Keb}. If $\beta'$ also vanishes at the endpoints, a second integration by parts yields
\begin{equation}
b_n=-\frac{L}{4\pi^2n^2}\int_I\beta''(y)
\cos{\left(\frac{2\pi n}{L}\left(y+\frac{L}{2}\right)\right)}~dy,
\end{equation}
which proves \eqref{W21Keb}.
\end{proof}
Notice that the estimates for $b_n$ use only $\beta\in W^{1,1}(I)$ for the first integration by parts and $\beta\in W^{2,1}(I)$ with the stated endpoint conditions for the second while 
the operator-norm remainders use the stronger perturbation assumptions adopted above. Finally, \eqref{crshift} shows that an even velocity perturbation changes all mode energies by the same relative amount to first order. Adjusting the transfer time to the perturbed optical crossing time cancels this contribution. If $\beta=0$, the Luttinger parameter remains constant and the perturbed channel is still conformal. In that case, oscillator PWT occurs exactly at
\begin{equation}
T_\eta=\int_I\frac{dx}{v_\eta(x)}.
\end{equation}
Moreover, spatial variation in the Luttinger perturbation
produces the mode-dependent shifts $b_n$. Their Fourier decay controls the first-order sensitivity of the higher modes. In particular, \eqref{W11Keb} bounds the first-order transfer-error coefficient independently of $N$, while the additional endpoint conditions in \eqref{W21Keb} make the individual first-order phase errors decay as $1/n$. The constants in the $O_N(\eta^2)$ remainders may depend on $N$, so a bound for the full error uniform in bandwidth requires control of
the higher-order terms.

\subsection{Ultraviolet dispersion and transfer bandwidth}

The TLL description is based on a linearized low-energy dispersion. At higher energies, band curvature and nonlinear hydrodynamic terms modify the dynamics of one-dimensional quantum liquids~\cite{ImambekovGlazman2009,ImambekovSchmidtGlazman2012}. We first consider dispersive corrections described by a
reflection-symmetric quadratic Hamiltonian. The transfer errors below are evaluated on its first $N$ positive normal modes and on the associated Fock space with total occupation at most $M$. We also assume that these modes have reflection parity $p_n=(-1)^n$. Let $\{E_n^{(0)}\}_{n\geq 1}$ be a strictly increasing positive reference spectrum with oscillator PWT at $T_0$, so that
\begin{equation}\label{uvrs}
E_n^{(0)}T_0=\pi m_n.
\end{equation}
Here, the positive integers $m_n$ form a strictly increasing sequence, with $m_n$ even for even $n$ and odd for odd $n$.  Let us write the corrected energies as
\begin{equation}
\widetilde{E}_n=(1+\gamma)E_n^{(0)}+\Delta_n^{\mathrm{UV}},
\qquad 1+\gamma>0.
\end{equation}
The common rescaling of the mode energies leads to the adjusted transfer time $\widetilde{T}_0=T_0/(1+\gamma)$. At this time, the accumulated phase differs from the required
reflection phase by
\begin{equation}\label{uvepd}
\theta_n^{\mathrm{UV}}:=\widetilde{E}_n\widetilde{T}_0-\pi m_n
=\frac{T_0}{1+\gamma}\Delta_n^{\mathrm{UV}}.
\end{equation}
Thus, $e^{-i\widetilde{E}_n\widetilde{T}_0}=p_n e^{-i\theta_n^{\mathrm{UV}}}$. Notice that we use $\theta_n^{\mathrm{UV}}$ for this real phase difference, which may lie outside the principal interval used to define $\delta_n$ earlier. A bound on the dispersive corrections now gives a sufficient condition on the number of modes and total occupation for transfer with a prescribed accuracy. Let $n_{\mathrm{UV}}>0$ denote the characteristic mode-index scale entering this bound.
\begin{corollary}
Suppose that, throughout the band $1\leq n\leq N$, the following estimate holds
\begin{equation}\label{uvrda}
\left|\frac{\Delta_n^{\mathrm{UV}}}{E_n^{(0)}}\right|
\leq A\left(\frac{n}{n_{\mathrm{UV}}}\right)^p
\end{equation}
with $A>0$ and $p>0$. Then, the one-particle transfer error satisfies
\begin{equation}\label{uvopeb}
\varepsilon_N^{(1)}(\widetilde{T}_0)\leq\min\left\{2,\frac{\pi A}{1+\gamma}m_N\left(\frac{N}{n_{\mathrm{UV}}}\right)^p\right\}.
\end{equation}
For states with total occupation bounded by $M$, where $M\geq 1$, the corresponding Fock-space transfer error satisfies
\begin{equation}\label{uvfeb}
\varepsilon_{N,M}^{\mathrm{F}}(\widetilde{T}_0)\leq
\min\left\{2,\frac{\pi AM}{1+\gamma}m_N\left(\frac{N}{n_{\mathrm{UV}}}\right)^p\right\}.
\end{equation}
For a conformal reference channel at its first oscillator-transfer time, $m_n=n$. In this case, an error tolerance
$0<\epsilon_{\mathrm{tar}}<2$ is guaranteed whenever
\begin{equation}\label{uvbw}
N\leq
\left[\frac{(1+\gamma)\epsilon_{\mathrm{tar}}}{\pi A M} n_{\mathrm{UV}}^p\right]^{1/(p+1)},
\end{equation}
provided \eqref{uvrda} applies throughout the selected band.
\end{corollary}
\begin{proof}
If we combine \eqref{uvrs}, \eqref{uvepd} and \eqref{uvrda}, we obtain
\begin{equation}
|\theta_n^{\mathrm{UV}}|\leq\frac{\pi A}{1+\gamma}
m_n\left(\frac{n}{n_{\mathrm{UV}}}\right)^p.
\end{equation}
Since $m_n$ is increasing, the right-hand side is bounded
by its value at $n=N$. The one-particle error is
$\max_{1\leq n\leq N}|e^{-i\theta_n^{\mathrm{UV}}}-1|$,
so $|e^{-ix}-1|\leq\min\{2,|x|\}$ proves \eqref{uvopeb}.
On an occupation state, the phase difference is $\sum_{n=1}^N k_n\theta_n^{\mathrm{UV}}$. Applying the same inequality with $\sum_n k_n\leq M$, as in Theorem~\ref{tfbne}, gives \eqref{uvfeb}. Finally, if we set $m_N=N$ and solve for $N$, we get \eqref{uvbw}.
\end{proof}
Equation~\eqref{uvbw} relates the usable transfer bandwidth to the allowed occupation and error tolerance within the
quadratic description. The parameters $A$, $p$ and $n_{\mathrm{UV}}$ must be determined from the microscopic model or a controlled higher-gradient approximation. The resulting accuracy bound applies over the range where \eqref{uvrda} is justified. Notice that interactions between modes can also cause scattering, generate occupation-dependent phases and populate modes outside the hydrodynamic band. Assessing transfer in this setting requires the full many-body evolution. Once the low-energy states have been embedded in the microscopic Hilbert space, let
$\mathsf{P}_{N,M}^{\mathrm{micro}}$ project onto the selected
initial subspace. The relevant error is then $\left\|
\left(U_{\mathrm{micro}}(T)-P_{\mathrm{micro}}\right)
\mathsf{P}_{N,M}^{\mathrm{micro}}\right\|$ where $P_{\mathrm{micro}}$ is spatial reflection in that Hilbert space. This norm also accounts for transitions out of the selected initial subspace. Evaluating it or the transfer fidelity of a given state requires information about the mode couplings in addition to the corrected energies.

\subsection{Finite-site benchmarks of rational transfer channels}\label{dlb}

\begin{figure}[t]
\centering
\includegraphics[width=\linewidth]{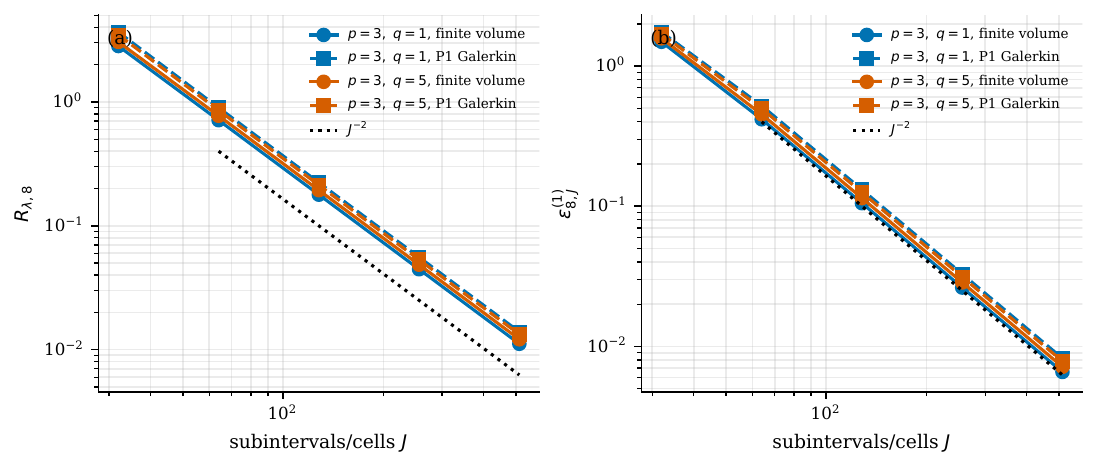}
\caption{Fixed-band convergence for $r=3$ and $r=3/5$ at
dimensionless time $3\pi$. The squared-frequency residual and one-particle transfer error are evaluated on the first eight nonzero modes using the cell-centered finite-volume and consistent-mass piecewise-linear Galerkin schemes. The reference lines indicate second-order convergence as the number of cells or elements increases.}
\label{Figure_rc}
\end{figure}

\begin{figure}[t]
\centering
\includegraphics[width=\linewidth]{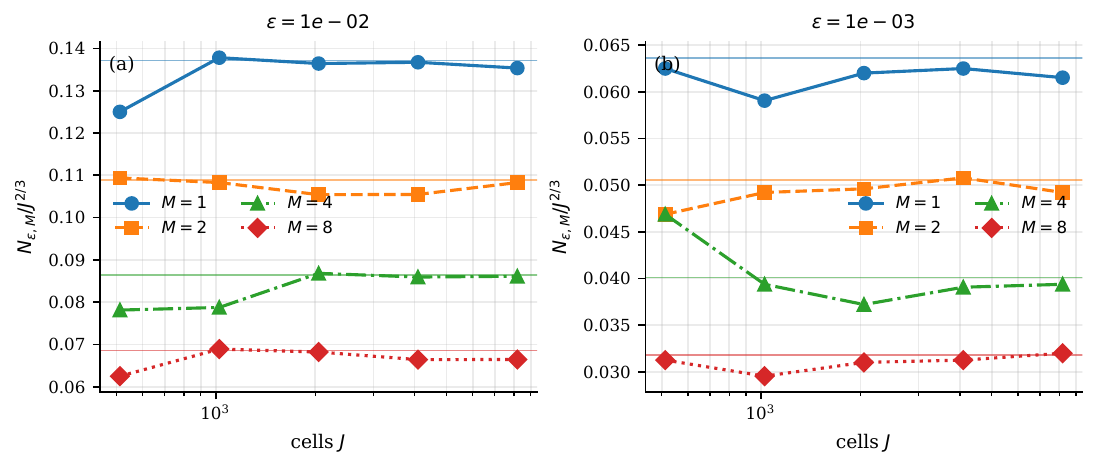}
\caption{Usable transfer bandwidth of the $r=3/5$ finite-volume chain at dimensionless time $3\pi$. The scaled bandwidth $N_{\epsilon,M}(J)/J^{2/3}$ is shown for total occupation bounds $M=1,2,4,8$ and tolerances $\epsilon=10^{-2}$ and $10^{-3}$. Horizontal lines give the small-tolerance prediction in \eqref{prbl}. For these tolerances, the inverse-sine correction changes the predicted coefficient by less than $4.2\times10^{-6}$ in relative terms. The numerical results support the predicted growth of the usable band.}
\label{fig_rb}
\end{figure}

\begin{figure}[t]
\centering
\includegraphics[width=\linewidth]{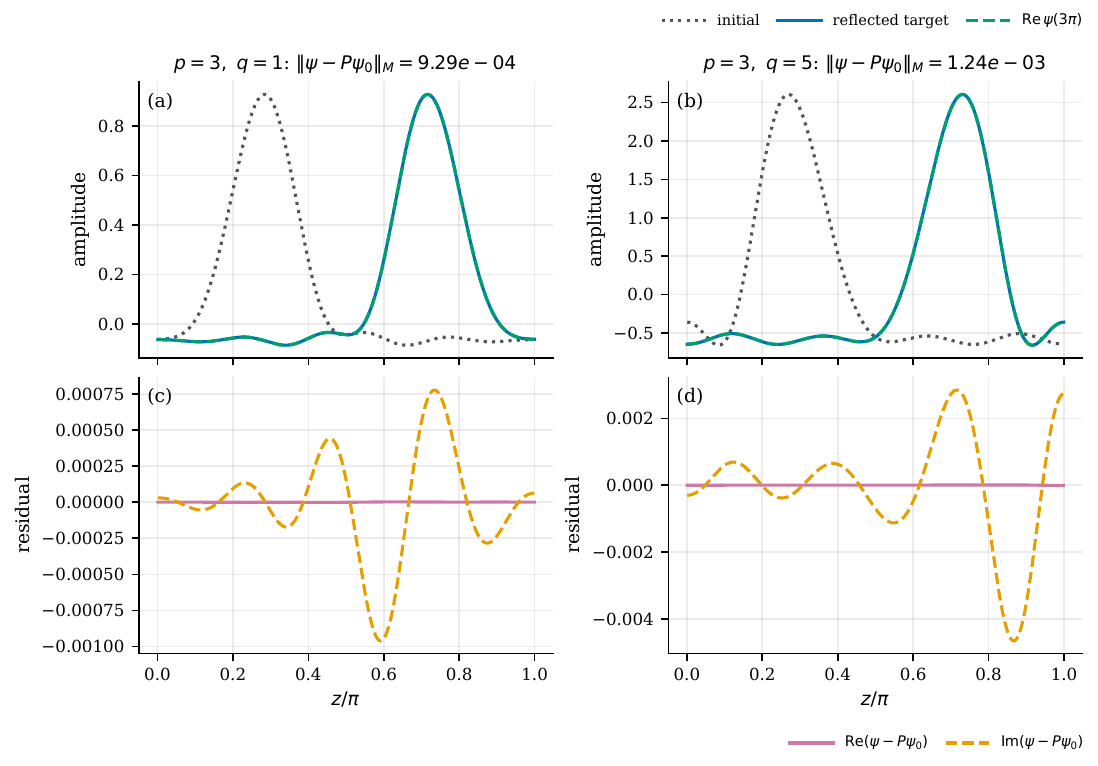}
\caption{Transfer of a Gaussian trial profile projected onto the first eight nonzero modes of the $J=512$ finite-volume chains. The two designs are $r=3$ and $r=3/5$, evaluated at dimensionless time $3\pi$. The upper panels show the initial profile, its reflected image and the real part of the evolved profile. The lower panels show the
real and imaginary parts of the evolved profile minus the reflected initial profile.}
\label{Figure_rtr}
\end{figure}

\begin{figure}[t]
\centering
\includegraphics[width=\linewidth]{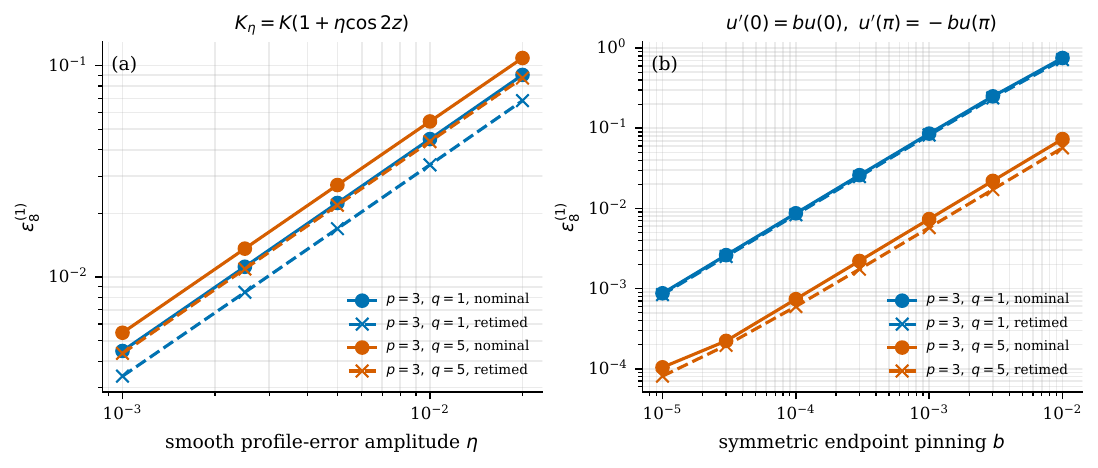}
\caption{Sensitivity of the $r=3$ and $r=3/5$ designs to two symmetric perturbations. (a) A bulk change $K_\eta=K_r(1+\eta\cos{2z})$ at fixed velocity. (b) Endpoint pinning with
$u'(0)=bu(0)$ and $u'(\pi)=-bu(\pi)$. Solid curves show the first-eight-mode errors at the nominal transfer time $3\pi$ and dashed curves use an optimized transfer time. The $r=3/5$ profile has a larger error under the tested bulk
perturbation and a smaller error under endpoint pinning.}
\label{Figure_rr}
\end{figure}

The explicit rational profiles allow us to test delayed oscillator transfer
in local harmonic chains. More precisely, we compare $r=3$ and $r=3/5$, corresponding to $(p,q)=(3,1)$ and $(3,5)$, which share the shortest continuum transfer time $3L/v_0$. The discretization is defined for a general admissible
$r=p/q$. Using the dimensionless coordinate $z\in[0,\pi]$ and measuring
time in units of $L/(\pi v_0)$, the continuum frequencies and shortest
transfer time are $\omega_0=0$, $\omega_1=q/p$, $\omega_n=n$ with $n\geq 2$ and $\tau_p=p\pi$. Moreover, we partition the interval into $J$ cells of width $h=\pi/J$ with centers $z_j=(j-1/2)h$. The masses and spring constants are $M_j=hK_r(z_j)$ with $j=1,\ldots,J$ and $\kappa_{j+1/2}=K_r(jh)/h$ with $j=1,\ldots,J-1$, respectively. Moreover, they determine the Hamiltonian of a local harmonic chain, that is
\begin{equation}
H_J=\frac{1}{2}\sum_{j=1}^{J}\frac{P_j^2}{M_j}+\frac{1}{2}\sum_{j=1}^{J-1}\kappa_{j+1/2}(Q_{j+1}-Q_j)^2.
\end{equation}
Notice that the missing exterior springs implement free endpoints. Furthermore, the normal modes solve $S_J f_{n,J}=\omega_{n,J}^2 M_J f_{n,J}$ where $M_J=\operatorname{diag}(M_j)$ and
\begin{equation}
f^\mathsf{T}S_J f=\sum_{j=1}^{J-1}\kappa_{j+1/2}(f_{j+1}-f_j)^2.
\end{equation}
This cell-centered finite-volume discretization of $-K_r^{-1}(K_ru')'$ preserves reflection symmetry and the constant zero mode. We order its frequencies as
$0=\omega_{0,J}<\omega_{1,J}<\cdots<\omega_{J-1,J}$. The mode with index $n$ has reflection parity $(-1)^n$. For $1\leq N\leq J-1$, we measure the spectral and transfer errors by
\begin{equation}
R_{\Lambda,N}(J):=\max_{0\leq n\leq N}\left|\omega_{n,J}^2-\omega_n^2\right|,\qquad
\varepsilon_{N,J}^{(1)}:=\max_{1\leq n\leq N}\left|e^{-i\pi p\omega_{n,J}}-(-1)^n\right|.
\end{equation}
The dimensionless residual $R_{\Lambda,N}$ corresponds to a physical squared-energy error $(\pi v_0/L)^2 R_{\Lambda,N}$. To evaluate transfer for several oscillator quanta, we introduce the principal phase defects $\delta_{n,J}\in(-\pi,\pi]$ through
$e^{-ip\pi\omega_{n,J}}=(-1)^n e^{-i\delta_{n,J}}$. For states in the first $N$ nonzero modes with total occupation at most $M$, the oscillator transfer error is
\begin{equation}\label{ddfe}
\varepsilon_{N,M,J}^{\mathrm{F}}:=\max_{\substack{k_1,\ldots,k_N\in\mathbb{N}_0\\
\sum_{n=1}^N k_n\leq M}}\left|\exp\left(-i\sum_{n=1}^N k_n\delta_{n,J}\right)-1\right|.
\end{equation}
This expression follows from Theorem~\ref{tfbne} applied to the nonzero modes of the finite chain after removing the state-independent vacuum phase. We next ask how many consecutive low-frequency modes satisfy a prescribed error tolerance $0<\epsilon<2$. To this aim, we define $N_\epsilon(J):=\max\left\{N\in\{0,1,\ldots,J-1\}:\varepsilon_{N,J}^{(1)}\leq\epsilon\right\}$ and for an occupation cap $M\in\mathbb N$ we introduce the quantity $N_{\epsilon,M}(J):=\max\left\{N\in\{0,1,\ldots,J-1\}:\varepsilon_{N,M,J}^{\mathrm{F}}\leq\epsilon\right\}$. Moreover, We also set both transfer errors to zero at $N=0$. Thus, $N_\epsilon(J)$ counts the consecutive nonzero modes that satisfy the one-particle tolerance, while $N_{\epsilon,M}(J)$ imposes the same tolerance for all states with total occupation at most $M$ in that band. These quantities measure how the usable band grows as the number of sites increases. For fixed mode index, the smooth positive profile permits a second-order variational approximation. Midpoint quadrature gives
errors of order $h^2$ in the mass and interior energy forms evaluated
on smooth modes while the Neumann conditions make the omitted endpoint half-cell energies of order $h^3$. Together with stability and the corresponding spectral approximation estimates~\cite{Keller1965,BabuskaOsborn1989,BanerjeeOsborn1989},
these give, for each fixed positive $n$
\begin{equation}\label{dler}
|\omega_{n,J}-\omega_n|=O_n(J^{-2}).
\end{equation}
Hence, for fixed $N$ and $M$, we have
\begin{equation}\label{dler2}
R_{\Lambda,N}(J)=O_N(J^{-2}),\qquad
\varepsilon_{N,J}^{(1)}=O_N(J^{-2}),\qquad
\varepsilon_{N,M,J}^{\mathrm F}=O_{N,M}(J^{-2}).
\end{equation}
The constants in \eqref{dler2} may depend on the selected band and occupation bound. Table~\ref{tab_dld} reports the results for $r=3$, whose Luttinger profile has the larger ratio $K_{\max}/K_{\min}$.  The Fock-space errors are obtained by evaluating the finite maximum in \eqref{ddfe}. Moreover, the growth of the usable band requires estimates in which the mode index increases with $J$. For the homogeneous chain, the exact sine dispersion yields the bandwidth asymptotics proved in Theorem~\ref{hrba}. For the engineered profiles, the leading part of the centered
discretization suggests the formal expansion
\begin{equation}
\omega_{n,J}= n-\frac{\pi^2 n^3}{24J^2}+O_{K_r}\left(\frac{n}{J^2}+\frac{n^5}{J^4}\right)
\end{equation}
in the range $1\ll n\ll J$. If the expansion holds uniformly up to $n=O(J^{2/3})$, the real phase difference
$\widetilde{\delta}_{n,J}:=\pi p(\omega_{n,J}-n)$ has leading term $\widetilde{\delta}_{n,J}\simeq -\pi^3 pn^3/(24J^2)$. Applying the relation between phase differences and transfer errors used in the proof of Theorem~\ref{hrba} gives the prediction
\begin{equation}\label{prbl}
N_{\epsilon,M}^{\mathrm{pred}}(J)=\sqrt[3]{\frac{48\arcsin(\epsilon/2)}{\pi^3 pM}}J^{2/3}\sim
\sqrt[3]{\frac{24\epsilon}{\pi^3 pM}}J^{2/3}
\end{equation}
as $\epsilon\downarrow 0$. The coefficient in \eqref{prbl} is specified by $p$, $M$ and $\epsilon$ with no fitted parameters. For the engineered chains, this law remains a prediction whose proof requires a spectral estimate uniform up to $n=O(J^{2/3})$. Notice that such an estimate would extend the fixed-mode convergence in \eqref{dler} to the growing band. For $p=3$ and $M=1$, the fixed-tolerance prefactors are $0.1371558$ for $\epsilon=10^{-2}$ and $0.0636620$ for $\epsilon=10^{-3}$. At $J=8192$, the $r=3/5$ chain has usable bandwidths $N_{\epsilon,M}=55,44,35,27$ for occupation bounds $M=1,2,4,8$, respectively, when $\epsilon=10^{-2}$. For $\epsilon=10^{-3}$, the corresponding bandwidths are $25,20,16,13$. These values agree closely with the prediction. For each reported bandwidth, the calculation also checks that adding the next mode exceeds the prescribed error tolerance. Hence, the numerical agreement supports the predicted scaling for the engineered profiles. 
Figure~\ref{fig_rb} shows the scaled bandwidth for $r=3/5$.
\begin{table*}
\caption{Finite-volume results for $r=3$ at dimensionless time $3\pi$. The spectral residual and transfer errors use the first eight nonzero modes. The Fock-space columns impose total occupation bounds of two and four. The last two columns give the number of consecutive nonzero modes satisfying the indicated one-particle error tolerance. A zero means that the first nonzero mode already exceeds that tolerance. Corresponding data for $r=3/5$ are supplied in the repository.}
\label{tab_dld}
\centering
\small
\begin{tabular}{c|cccc|cc}
  $J$ & $R_{\Lambda,8}$ & $\varepsilon_{8,J}^{(1)}$
      & $\varepsilon_{8,2,J}^{\mathrm F}$ & $\varepsilon_{8,4,J}^{\mathrm F}$
      & $N_{10^{-2}}(J)$ & $N_{10^{-3}}(J)$\\
  \hline
    64 & $7.116\times10^{-1}$ & $4.172\times10^{-1}$ & $8.161\times10^{-1}$ & $1.490$                 & 3  & 0\\
   128 & $1.783\times10^{-1}$ & $1.050\times10^{-1}$ & $2.098\times10^{-1}$ & $4.172\times10^{-1}$  & 4  & 0\\
   256 & $4.459\times10^{-2}$ & $2.627\times10^{-2}$ & $5.253\times10^{-2}$ & $1.050\times10^{-1}$  & 6  & 3\\
   512 & $1.115\times10^{-2}$ & $6.567\times10^{-3}$ & $1.313\times10^{-2}$ & $2.627\times10^{-2}$  & 9  & 4\\
  1024 & $2.787\times10^{-3}$ & $1.642\times10^{-3}$ & $3.283\times10^{-3}$ & $6.567\times10^{-3}$  & 14 & 6\\
  2048 & $6.968\times10^{-4}$ & $4.104\times10^{-4}$ & $8.209\times10^{-4}$ & $1.642\times10^{-3}$  & 22 & 10\\
  4096 & $1.742\times10^{-4}$ & $1.026\times10^{-4}$ & $2.052\times10^{-4}$ & $4.104\times10^{-4}$  & 35 & 16\\
  8192 & $4.355\times10^{-5}$ & $2.565\times10^{-5}$ & $5.130\times10^{-5}$ & $1.026\times10^{-4}$  & 55 & 25
\end{tabular}
\end{table*}
We check fixed-band convergence independently using a nodal
piecewise-linear Galerkin discretization. For this comparison, let $\{\varphi_j\}_{j=0}^{J}$ be the continuous, piecewise-linear basis functions associated with the nodes $z_j=jh$, defined by $\varphi_j(z_i)=\delta_{ij}$ and linear interpolation between adjacent nodes. The generalized matrices are
\begin{equation}
(S_J^{\mathrm{G}})_{ij}=\int_0^\pi K_r(z)\varphi_i'(z)\varphi_j'(z)~dz,\qquad
(M_J^{\mathrm{G}})_{ij}=\int_0^\pi K_r(z)\varphi_i(z)\varphi_j(z)~dz.
\end{equation}
The weak formulation incorporates the Neumann conditions naturally. We evaluate the element integrals using four-point Gauss--Legendre quadrature. Figure~\ref{Figure_rc} compares the two schemes for both profiles. Here, $J$ counts cells or elements, giving $J$ degrees of freedom for the finite-volume scheme and $J+1$ for the Galerkin scheme. The numerical values are provided in Appendix~\ref{ANC}. Both discretizations exhibit second-order convergence of the fixed-band errors. The fitted slopes over $J=128,256,512$ lie between $-2.0012$ and $-1.9992$ across both profiles, both solvers, and both fixed-band diagnostics.
At $J=512$, a Gaussian trial profile projected onto the first eight nonzero modes and propagated to $3\pi$ gives mass-norm residuals $9.29\times10^{-4}$ for $r=3$ and $1.24\times10^{-3}$ for $r=3/5$. Figure~\ref{Figure_rtr} compares the evolved and reflected profiles and displays the real and imaginary parts of their difference. The mass norm is $\|f\|_{M_J}=\sqrt{f^\dagger M_J f}$.
The relative sensitivity of the two designs depends on the form of the perturbation. Figure~\ref{Figure_rr} compares a smooth symmetric change in $K$ with symmetric endpoint pinning. For the bulk perturbation $K_\eta=K_r(1+\eta\cos{2z})$ at fixed velocity and $\eta=10^{-2}$, the first-eight-mode errors at the nominal and optimized transfer times are
\begin{equation}
 \begin{array}{c|cc}
 r & \varepsilon_8^{(1)} & \varepsilon_{8,\mathrm{ret}}^{(1)}\\ \hline
 3   &0.0449320&0.0339548\\
 3/5 &0.0544391&0.0437241
 \end{array}.
\end{equation}
Notice that the $r=3/5$ profile has the larger error for this bulk perturbation. The ordering reverses for symmetric endpoint pinning $u'(0)=bu(0)$, $u'(\pi)=-bu(\pi)$ with $b=10^{-3}$. The corresponding errors are $0.0858402$ ($0.0822926$ retimed) for $r=3$ and $0.00736183$
($0.00575717$ retimed) for $r=3/5$. The $r=3/5$ design therefore has an error approximately one order of magnitude smaller for this boundary perturbation. The shortest spatial variation scale, defined by $\ell_K^{-1}=\max_y|\partial_y\log K|$, increases from $\ell_K/L=0.0344581$ for $r=3$ to $\ell_K/L=0.0861451$ for $r=3/5$. This larger scale makes it easier to satisfy the separation between the microscopic cutoff length and the profile variation required for the local TLL description.
The accompanying script
\texttt{codes/pwt\_revised\_reproducibility.py} generates the spectra, transfer errors, bandwidths and figures. It also checks reflection parity and the zero-mode residual and evaluates the occupation-constrained maxima in \eqref{ddfe}. 

These computations provide finite-site evidence for delayed transfer in quadratic oscillator systems. Each site has a continuous coordinate, so the chain has an infinite-dimensional Hilbert space. Its constant mode describes a noncompact free particle and the transfer diagnostics
above concern the nonzero oscillator modes. Extending the benchmark to global transfer in an interacting compact-field model would require incorporating the integer-charge rotor of Sec.~\ref{pwt_notions} together with the effects of interactions, including scattering and transitions to modes outside the selected band.

\subsection{Transfer errors in a homogeneous harmonic chain}

The homogeneous harmonic chain provides an explicit setting in which to derive finite-band error bounds and prove the bandwidth scaling suggested by the engineered channels. Its exact dispersion relation allows us to control the error as both the number of sites and the number of included modes increase. To this aim, let us consider $J$ canonical pairs $(\phi_j,\Pi_j)$ at the cell centers $x_j=(j-1/2)a$ of an interval of length $L=Ja$. The Hamiltonian is
\begin{equation}
H_{\mathrm{lat}}=\frac{1}{2}\sum_{j=1}^{J}\Pi_j^2+\frac{v^2}{2a^2}\sum_{j=1}^{J-1}(\phi_{j+1}-\phi_j)^2.
\end{equation}
Notice that the absence of exterior springs gives free endpoints. In addition to the constant zero mode, the chain has $J-1$ nonzero normal modes
\begin{equation}\label{hcs}
f_n(j)=\sqrt{\frac{2}{J}}\cos{\left(\frac{\pi n}{J}\left(j-\frac{1}{2}\right)\right)},\qquad
\omega_n=\frac{2v}{a}\sin{\left(\frac{\pi n}{2J}\right)}.
\end{equation}
Here, $1\leq n\leq J-1$ and the modes are normalized in the
Euclidean inner product. Reflection acts through
$f_n(J+1-j)=(-1)^n f_n(j)$. As $J\to\infty$ at fixed $L$, each fixed-mode frequency approaches $\pi n v/L$. The continuum reference channel therefore has its first
oscillator-transfer time at $T_0=L/v$. We restrict the following transfer errors to the nonzero oscillator
modes and remove the state-independent vacuum phase. The sine dispersion then gives a bound valid for every finite $J$.
\begin{proposition}
At time $T_0$, the phase difference from the continuum value $\pi n$ is
\begin{equation}\label{hcud}
\widetilde{\delta}_n^{\mathrm{lat}}:=\omega_n T_0-\pi n
=2J\sin{\left(\frac{\pi n}{2J}\right)}-\pi n.
\end{equation}
For every $1\leq n\leq J-1$, it satisfies $|\widetilde{\delta}_n^{\mathrm{lat}}|\leq (\pi^3n^3)/(24J^2)$. Its representative $\delta_n^{\mathrm{lat}}$ in $(-\pi,\pi]$ has
$|\delta_n^{\mathrm{lat}}|\leq|\widetilde{\delta}_n^{\mathrm{lat}}|$. Moreover, for the first $N$ nonzero modes, these phase estimates yield the following bounds on the one-particle and occupation-constrained transfer errors
\begin{eqnarray}
\varepsilon_N^{(1)}(T_0)&\leq&\min\left\{2,\frac{\pi^3 N^3}{24J^2}\right\},\label{hcopb}\\
\varepsilon_{N,M}^{\mathrm{F}}(T_0)&\leq&\min\left\{2,\frac{\pi^3 MN^3}{24J^2}\right\},\label{hcfb}
\end{eqnarray}
where $1\leq N\leq J-1$ and $M$ is a positive integer
bounding the total occupation. For an error tolerance
$0<\epsilon_{\mathrm{tar}}<2$, a sufficient condition for
$\varepsilon_{N,M}^{\mathrm{F}}(T_0)\leq\epsilon_{\mathrm{tar}}$ is
\begin{equation}\label{hcbw}
N\leq\sqrt[3]{\frac{24\epsilon_{\mathrm{tar}}}{\pi^3 M}}J^{2/3}.
\end{equation}
Furthermore, as $n/J\to 0$, the phase difference has the expansion
\begin{equation}\label{hcda}
\widetilde{\delta}_n^{\mathrm{lat}}=-\frac{\pi^3 n^3}{24J^2}
+O\left(\frac{n^5}{J^4}\right).
\end{equation}
\end{proposition}
\begin{proof}
Substituting $T_0=Ja/v$ into \eqref{hcs} gives \eqref{hcud}.
With $x_n=\pi n/(2J)$, the inequality $0\leq x-\sin x\leq x^3/6$ for $x\geq 0$ yields $|\widetilde{\delta}_n^{\mathrm{lat}}|=2J(x_n-\sin{x_n})
\leq(\pi^3 n^3)/(24J^2)$. If we add an integer multiple of $2\pi$ to place the phase difference in $(-\pi,\pi]$, its complex exponential remain unchanged and cannot increase its absolute value. Proposition~\ref{opfbe} and Theorem~\ref{tfbne} therefore give \eqref{hcopb} and
\eqref{hcfb}. Solving the latter bound for $N$ gives \eqref{hcbw}, while Taylor expansion of the sine gives \eqref{hcda}.
\end{proof}
Since the leading phase correction grows as $n^3/J^2$, it  suggests a usable bandwidth proportional to $J^{2/3}$ at fixed occupation and error tolerance. The next result establishes this scaling and its coefficient
at every odd multiple of the continuum transfer time.
\begin{theorem}\label{hrba}
Let us fix an odd integer $p\geq 1$, a positive integer occupation bound $M$ and a tolerance $0<\epsilon<2$. At time $pT_0$, let $N_{\epsilon,M}^{\mathrm{hom}}(J;p)$ be the largest $N\in\{0,\ldots,J-1\}$ for which the exact oscillator transfer error on the first $N$ nonzero modes, with total occupation at most $M$, is at most $\epsilon$. The error at $N=0$ is defined to be zero. As the number of sites increases, the bandwidth satisfying the prescribed error tolerance obeys the asymptotic law
\begin{equation}\label{ehbl}
\lim_{J\to\infty}\frac{N_{\epsilon,M}^{\mathrm{hom}}(J;p)}{J^{2/3}}=\sqrt[3]{\frac{48\arcsin{(\epsilon/2)}}{\pi^3 pM}}.
\end{equation}
The coefficient governing this growth has the following expansion as $\epsilon\to 0$
\begin{equation}\label{hbse}
\sqrt[3]{\frac{48\arcsin{(\epsilon/2)}}{\pi^3 pM}}=
\sqrt[3]{\frac{24\epsilon}{\pi^3 pM}}\left[1+\frac{\epsilon^2}{72}+O(\epsilon^4)\right].
\end{equation}
\end{theorem}
\begin{proof}
Let us define $d_{n,J}:=\pi n-2J\sin{(\pi n/(2J))}$ with $d_{0,J}=0$. For fixed $J$, $d_{n,J}$ is positive and strictly increasing for $n\geq 1$. Since $p$ is odd, $e^{-i\pi p n}=(-1)^n$, so the phase difference relative to reflection at time $pT_0$ is $-p d_{n,J}$. Morever, we set $a_\epsilon=2\arcsin{(\epsilon/2)}\in(0,\pi)$. If $pM d_{N,J}\leq\pi$, every occupation state in the selected
subspace has phase magnitude at most $pM d_{N,J}$. On this interval, the error $2\sin{(x/2)}$ increases with the phase
magnitude $x$. Its maximum is attained by placing all $M$ quanta in mode $N$, leading to $\varepsilon_{N,M,J}^{\mathrm{hom}}=2\sin{(pMd_{N,J}/2)}$. 
Thus, the threshold index $N_*(J):=\max\left\{
N\in\{0,\ldots,J-1\}\mathrel{\big|}
pM d_{N,J}\leq a_\epsilon\right\}$ defines a band whose error satisfies the tolerance. For $n=O(J^{2/3})$, Taylor's theorem gives uniformly
\begin{equation}
d_{n,J}=\frac{\pi^3 n^3}{24J^2}\left[1+O\left(\frac{n^2}{J^2}\right)\right]=\frac{\pi^3 n^3}{24J^2}
\left[1+O\left(\frac{1}{J^{2/3}}\right)\right].
\end{equation}
If we evaluate this expression at a sufficiently large fixed multiple of $J^{2/3}$ and use monotonicity, we can show that $N_*(J)=O(J^{2/3})$. In particular, $N_*(J)<J-1$ for all sufficiently large $J$. In this range, the difference between consecutive values is $d_{n+1,J}-d_{n,J}=O(J^{-2/3})$. The definition of $N_*(J)$ therefore implies $pM d_{N_*,J}\leq a_\epsilon<pM d_{N_*+1,J}=a_\epsilon+O(J^{-2/3})$. Since $a_\epsilon<\pi$, the phase magnitude $pM d_{N_*+1,J}$ lies between $a_\epsilon$ and $\pi$ for sufficiently large $J$. The state with all $M$ quanta in mode $N_*+1$ then has
error greater than $\epsilon$. Every larger band contains this state and also exceeds the tolerance. Hence, 
$N_{\epsilon,M}^{\mathrm{hom}}(J;p)=N_*(J)$ for sufficiently large $J$. The same consecutive-index estimate gives $pM d_{N_*,J}\to a_\epsilon$. If we substitute the uniform expansion of $d_{n,J}$ and let $J\to\infty$ with $p$, $M$ and $\epsilon$ fixed, we obtain
\begin{equation}
\frac{\pi^3 pM}{24}\left(\frac{N_*(J)}{J^{2/3}}\right)^3
\longrightarrow a_\epsilon,
\end{equation}
which proves \eqref{ehbl}. Finally,
$2\arcsin{(\epsilon/2)}=\epsilon+\epsilon^3/24+O(\epsilon^5)$ gives \eqref{hbse}.
\end{proof}
Hence, the homogeneous chain establishes the $J^{2/3}$ bandwidth law directly from its exact dispersion relation and provides a reference for the numerical results in Sec.~\ref{dlb}, where the same transfer diagnostics are applied to chains approximating the engineered delayed
channels.

\section{Conclusions and outlook}\label{conclusions}

We have established a necessary and sufficient spectral classification of oscillator perfect wave transfer in regular reflection-symmetric massless TLLs with bounded Liouville potential. Exact transfer permits only finitely many deviations from the conformal ladder, subject to
strict spectral ordering and the arithmetic conditions imposed by reflection parity. Each admissible spectrum determines a unique even Liouville potential and symmetric Robin parameter. Its positive zero mode reconstructs the Luttinger profile up to an overall scale, while
an optical reparametrization supplies the freedom to vary the velocity. The Robin conditions arise from the physical zero-current boundaries under the Liouville transformation. Within the transfer class, imposing Neumann conditions also on the transformed modes forces the Luttinger parameter to be spatially constant. Moreover, transfer at the optical crossing time $L/v_0$ has the same consequence even when the induced Robin parameter is allowed to vary. The explicit one-defect family shows how this classification can guide profile design. The family is regular for $r>1/2$ and oscillator transfer occurs precisely at reduced rational parameters $r=p/q$ with positive odd numerator and denominator. The shortest transfer time is $pL/v_0$. The rational extension permits substantially weaker spatial variation at the same transfer delay. Furthermore, the sequence $r=p/(p+2)$ approaches a constant profile smoothly as odd $p\to\infty$, while its shortest transfer time diverges. Along such a  sequence, exact nonconformal transfer is achieved with progressively weaker spatial variation of the Luttinger parameter at the cost of a longer transfer time. The compact zero mode determines whether oscillator transfer extends to coherent transfer across charge sectors. Oscillator and neutral transfer coincide in the single-charge model considered here. For the integer charge lattice and canonical scalar lift of reflection, global transfer additionally requires $T/(4W)\in\mathbb{Z}$. For the chosen reflection lift, scaling the Luttinger parameter allows the charge sectors to acquire a common phase while preserving the oscillator spectrum. This phase matching connects the spectral construction to the observable algebra used to encode quantum information and determines whether transfer extends to coherent superpositions of different charges. The finite-site calculations connect the continuum construction to quadratic lattice dynamics. Two independent discretizations exhibit second-order convergence on fixed mode bands, while the occupation-
and charge-constrained estimates quantify how phase errors affect many-body transfer. For the homogeneous chain, the exact dispersion relation yields a bandwidth proportional to $J^{2/3}$ at fixed occupation bound and error tolerance. The engineered chains show numerical agreement with the corresponding prediction in \eqref{prbl}. Their sensitivity also depends on the perturbation. The $r=3/5$ design has a smaller error under the tested endpoint pinning and a larger error under the tested smooth bulk perturbation.
A smaller ratio $K_{\max}/K_{\min}$ therefore provides no general ordering of transfer accuracy.

These results suggest concrete directions for further work. For the engineered chains, a spectral approximation uniform up to mode indices of order $J^{2/3}$ would establish the predicted bandwidth coefficient and determine how the spatial profile enters the corrections. The general realization theorem also provides a basis for designing profiles with multiple spectral defects at a prescribed transfer time. An optimization based on the error estimates could then ask which admissible spectra minimize sensitivity to a specified profile or boundary perturbation,
subject to constraints on $K_{\max}/K_{\min}$ and the shortest spatial variation scale. Such a calculation would quantify the tradeoffs between delay, spatial resolution, and transfer accuracy. A further step is to identify a microscopic interacting model whose low-energy parameters approximate the constructed profiles and to evaluate transfer for a specified family of initial states. The quadratic benchmarks provide reference dynamics for separating dispersive phase errors from interaction-induced scattering and transitions outside the selected band. If the microscopic model also accounts for the compact integer-charge degree of freedom, the calculation could test whether an adjustment of the oscillator transfer time preserves the relative phases between charge sectors. For multicomponent liquids or momentum--winding charge lattices,
the corresponding question is which quadratic charge forms admit simultaneous rephasing at an oscillator-transfer time.
This would extend \eqref{glc} and determine which charge
encodings support global transfer in those theories.

\subsection*{Funding}

The authors declare that no specific funding was received for this work.

\subsection*{Code and data availability}

The code and data used to obtain the numerical results and generate the figures are publicly available at
\url{https://github.com/dutykh/perfect-wave-transfer-tll/}. The repository also provides the software dependencies and instructions for reproducing the calculations and compiling the manuscript.

\subsection*{Conflict of interest statement}

On behalf of all authors, the corresponding author states that there is no conflict of interest.

\appendix

\section{One-defect reconstruction for \texorpdfstring{$r>1$}{r>1}}
\label{app_rod}

We derive the Darboux-Crum transformation underlying the profiles
of Theorem~\ref{rodf} in the restricted range $r=c>1$. Their regularity and transfer criterion are established in the main text. The calculation here gives the transformation operator and mode functions used in the endpoint analysis of Appendix~\ref{app_cc}, followed by additional endpoint values and their large-$c$ behavior. We use the functions $D_c$, $N_c$, and $\vartheta_c$ from Theorem~\ref{rodf} with the dimensionless coordinate $z=\pi(y-a)/L\in[0,\pi]$ and $a_c=\pi/(2c)$. Moreover, primes on functions of $z$ denote derivatives with respect to $z$. The free solutions $f(z)=\cos{z}$ and $g_c(z)=\cos{\vartheta_c(z)}$ 
have spectral parameters $1$ and $c^{-2}$, respectively. Their Wronskian is $W(f,g_c)=fg_c'-f'g_c=D_c/c$. The positivity of $D_c$ established in the main text ensures that this Wronskian has no zeros on the closed interval. Hence, the second-order Darboux-Crum operator~\cite{Crum1955}
\begin{equation}\label{Crum_i}
\mathcal{Q}_c[\varphi]:=\frac{W(f,g_c,\varphi)}{W(f,g_c)}
\end{equation}
has smooth coefficients and satisfies
\begin{equation}
\left(-\frac{d^2}{dz^2}+\widehat{V}_c\right)\mathcal Q_c=\mathcal{Q}_c\left(-\frac{d^2}{dz^2}\right),
\end{equation}
where
\begin{equation}
\widehat{V}_c=-2\frac{d^2}{dz^2}\log{W(f,g_c)}=-2\frac{d^2}{dz^2}\log{D_c}.
\end{equation}
This recovers the potential in \eqref{rational_family_fields}
directly from the two seed solutions. The endpoint calculation in Appendix~\ref{app_cc} establishes the spectrum in \eqref{rational_one_defect_spectrum} with $r=c$. 
The corresponding zero-energy ground state follows by applying the Darboux-Crum operator to the constant free solution $\mathcal{Q}_c[1]=N_c/D_c=s_c$. The intertwining identity then yields $-s_c''+\widehat{V}_c s_c=0$ with $\widehat{V}_c=s_c''/s_c$. Hence, the transformed positive zero mode reconstructs the
Luttinger profile through \eqref{rational_family_fields}. The transformed eigenfunction corresponding to the eigenvalue $c^{-2}$ is proportional to $f/W(f,g_c)$. Normalizing its value at $z=0$ to one gives
\begin{equation}\label{closed_fsm}
\psi_{1,c}(z)=\frac{\sin{a_c}\cos{z}}{D_c(z)}.
\end{equation}
This function is odd about the midpoint and has a single zero. For $n\geq 2$, the remaining eigenfunctions are proportional
to $\mathcal{Q}_c[\cos{(nz)}]$. Moreover, they satisfy the same Robin
conditions and have parity $(-1)^n$. Evaluation of the explicit zero mode gives $s_c(0)=s_c(\pi)=1$ and $s_c(\pi/2)=c^{-2}$. Its logarithmic derivative at the left endpoint is
\begin{equation}
\widehat{\rho}_c=\frac{s_c'(0)}{s_c(0)}=\frac{1-c^2}{c\tan{a_c}},\qquad
\rho_c=\frac{\pi}{L}\widehat{\rho}_c.
\end{equation}
For the dimensionless potential, direct evaluation yields
\begin{equation}
\widehat{V}_c(\pi/2)=2\left(1-\frac{1}{c^2}\right),\qquad
\widehat{V}_c(0)=\frac{2(c^2-1)(c^2\cos^2{a_c}-1)}{c^2\sin^2{a_c}}.
\end{equation}
At $c=3$, these potential values are $16/9$ and $368/9$, respectively, while $\widehat{\rho}_3=-8\sqrt{3}/3$. As $c\to\infty$, the endpoint parameter satisfies $|\widehat{\rho}_c|\sim 2c^2/\pi$. Since $K_c'(a)/K_c(a)=2\rho_c$, with this derivative taken
with respect to $y$, the logarithmic endpoint slope grows
quadratically with $c$. For the odd-integer transfer subfamily,
this gives the dependence of the endpoint slope on the delay.

\section{Darboux-Crum characteristic function}
\label{app_cc}

We derive the characteristic function for the regular family of
Theorem~\ref{rodf} by evaluating the boundary conditions after the
Darboux-Crum transformation. Here, $c>1/2$ denotes the parameter called $r$ in that theorem, so the calculation also covers the range $1/2<c<1$ used in the main text. We work on $z\in[0,\pi]$, with primes denoting derivatives with respect to $z$ and set $\mu:=c^{-2}$. The seed functions $f$ and $g$ are those defined in Appendix~\ref{app_rod} with $g=g_c$ and $W_c=W(f,g)=D_c/c$. They satisfy $f''=-f$ and $g''=-\mu g$.
The positivity of $D_c$ proved in Theorem~\ref{rodf} ensures that $W_c$ is nonzero throughout this parameter range. To evaluate the transformed solutions at the endpoints, we define $h:=W_c'/W_c$ and $R:=(f'g''-f''g')/W_c$. Expansion of the Wronskian in \eqref{Crum_i} along its last
column gives $\mathcal{Q}_c[\chi]=\chi''-h\chi'+R\chi$. For a free solution satisfying $-\chi''=\lambda\chi$, the transformation and its derivative therefore take the forms
\begin{eqnarray}
\mathcal{Q}_c[\chi]&=&(R-\lambda)\chi-h\chi',\label{Crum_aQs}\\
\partial_z\mathcal{Q}_c[\chi]&=&(-\lambda-h'+R)\chi'+(\lambda h+R')\chi.
\label{Crum_aQps}
\end{eqnarray}
The seed functions give $W_c(0)=W_c(\pi)=c^{-1}\sin{a_c}$ where $a_c=\pi/(2c)$ as in Theorem~\ref{rodf}. Their differential equations also imply $W_c'=(1-\mu)fg$ and $(f'g''-f''g')'=(1-\mu)f'g'$. If we use these identities and $f'(0)=f'(\pi)=0$, we obtain the endpoint values $h(0)=-\widehat{\rho}_c$, $h(\pi)=\widehat{\rho}_c$, $R(0)=1$, $R(\pi)=1$, $h'(0)=1-\mu-\widehat{\rho}_c^{2}$, $h'(\pi)=1-\mu-\widehat{\rho}_c^{2}$, $R'(0)=\widehat{\rho}_c$ and $R'(\pi)=-\widehat{\rho}_c$ with $\widehat{\rho}_c=(\mu-1)c\cot{a_c}$, in agreement with \eqref{rational_family_endpoint_data}. We now apply these formulas to $\varphi(z,\lambda)=\cos{(kz)}$ where $k=\sqrt{\lambda}$. At $z=0$, the values $\varphi=1$ and $\partial_z\varphi=0$ give $\mathcal{Q}_c[\varphi](0)=1-\lambda$ and $\partial_z\mathcal{Q}_c[\varphi](0)=\widehat{\rho}_c(1-\lambda)$. Hence, the solution
\begin{equation}
u_c(z,\lambda):=\frac{\mathcal{Q}_c[\varphi](z)}{1-\lambda}
\end{equation}
satisfies the normalization $u_c(0,\lambda)=1$ and the left
Robin condition $\partial_z u_c(0,\lambda)=\widehat{\rho}_c u_c(0,\lambda)$. Although this expression contains $1-\lambda$ in the denominator, $u_c$ extends analytically through $\lambda=1$. For each fixed $z$, the function $\cos{(\sqrt{\lambda} z)}$ is entire in $\lambda$. At $\lambda=1$, it equals the seed $f$, which is annihilated by $\mathcal{Q}_c$, so the numerator vanishes. Division by $1-\lambda$ therefore gives a removable singularity and the endpoint normalization extends to that value. At the right endpoint, we find  $\varphi(\pi,\lambda)=\cos{(\pi k)}$ and $\partial_z\varphi(\pi,\lambda)=-k\sin{(\pi k)}$. Substituting into \eqref{Crum_aQs} and \eqref{Crum_aQps} yields $\mathcal{Q}_c[\varphi](\pi)=(1-\lambda)\cos{(\pi k)}+\widehat{\rho}_c k\sin{(\pi k)}$ and $\partial_z\mathcal{Q}_c[\varphi](\pi)=(\lambda-\mu-\widehat{\rho}_c^{2})k\sin{(\pi k)}+\widehat{\rho}_c(\lambda-1)\cos{(\pi k)}$. In the combination defining the right Robin condition, the cosine terms cancel, as do the terms proportional to $\widehat{\rho}_c^{2}$. Thus, we obtain $\partial_z\mathcal{Q}_c[\varphi](\pi)+\widehat{\rho}_c\mathcal{Q}_c[\varphi](\pi)=(\lambda-\mu)k\sin{(\pi k)}$. After normalization at the left endpoint, the right boundary condition is determined by
\begin{equation}
\partial_z u_c(\pi,\lambda)+\widehat{\rho}_c u_c(\pi,\lambda)=\frac{\lambda-\mu}{1-\lambda}
\sqrt{\lambda}\sin{(\pi\sqrt{\lambda})}.
\end{equation}
With the convention
\begin{equation}\label{Omega_closed}
\Omega_c(\lambda):=-\left[\partial_z u_c(\pi,\lambda)+\widehat{\rho}_c u_c(\pi,\lambda)\right],
\end{equation}
the characteristic function is
\begin{equation}\label{closed_fcf}
\Omega_c(\lambda)=\frac{\lambda-c^{-2}}{\lambda-1}
\sqrt{\lambda}\sin{(\pi\sqrt{\lambda})}.
\end{equation}
To identify its zeros, we first suppose $c\neq 1$. The function $\sqrt{\lambda}\sin{(\pi\sqrt{\lambda})}$ is entire in $\lambda$ and the apparent singularity at $\lambda=1$ has the finite value $\Omega_c(1)=-\pi(1-\mu)/2\neq 0$. The free eigenvalue at $\lambda=1$ is therefore absent from the transformed spectrum. Near the origin, we have 
$\Omega_c(\lambda)=\pi\mu\lambda+O(\lambda^2)$, so zero is
a simple eigenvalue. Since $0<\mu<4$ and $\mu\neq 1$, the factor $\lambda-\mu$ gives a simple zero at $\lambda=\mu$. The remaining positive zeros are $\lambda=n^2$ for $n\geq 2$. For $\lambda=-\nu^2<0$ and $\nu>0$, we find $\sqrt{\lambda}\sin{(\pi\sqrt{\lambda})}=-\nu\sinh{(\pi\nu)}\neq 0$. Since the factor $(\lambda-c^{-2})/(\lambda-1)$ is also finite and nonzero for $\lambda<0$, the characteristic function has no negative zeros. The corresponding operator therefore has no negative eigenvalues. For $c=1$, the rational factor cancels identically, and $\Omega_1(\lambda)=\sqrt{\lambda}\sin{(\pi\sqrt{\lambda})}$ gives the free Neumann spectrum. After the rescaling $\Lambda=(\pi/L)^2\lambda$, these zeros establish the spectrum in \eqref{rational_one_defect_spectrum}
throughout the regular family.

\section{Hald's theorem for finite spectral deformations}
\label{app_hald}

The inverse reconstruction in Theorems~\ref{chr} and \ref{rodf}
uses the following consequence of Hald's results~\cite{Hald1978}.
We state it for an arbitrary symmetric reference potential and
give the rescaling from $[0,\pi]$ to a general interval. Let $I=[a,b]$ with $L=b-a>0$, and consider the Robin problem~\eqref{hgrp} for a real potential $V\in L^1(I)$ and a real parameter $\rho$. Reflection symmetry on this interval means
\begin{equation}\label{Hald_ep}
V(y)=V(a+b-y)
\end{equation}
for almost every $y\in I$. Its eigenvalues are real and simple. Moreover, we denote them by $\Lambda_n(V,\rho)$, indexed from $n=0$ in increasing order. The following theorem establishes existence and uniqueness when a finite spectral change preserves strict ordering.
\begin{theorem}\label{Hfd}
Let $(\widetilde{V},\widetilde{\rho})$ be a symmetric pair of the form specified above with ordered spectrum $(\widetilde{\Lambda}_n)_{n\geq  0}$. We further suppose that a strictly increasing sequence of real numbers $(\Lambda_n)_{n\geq 0}$ satisfies $\Lambda_n=\widetilde{\Lambda}_n$ for every $n>N$ for some integer $N\geq 0$. Then, there exist a real parameter $\rho$ and a real potential
$V\in L^1(I)$ satisfying \eqref{Hald_ep} for which the spectrum of \eqref{hgrp} is exactly $(\Lambda_n)_{n\geq 0}$. The parameter is uniquely determined and any two symmetric potentials realizing this spectrum agree almost everywhere on $I$. Moreover, $V-\widetilde{V}$ agrees almost everywhere with a continuous function on $I$. In particular, if $\widetilde{V}$ has a continuous representative, then so does $V$.
\end{theorem}
\begin{proof}
To pass to Hald's interval, we set $\xi=\pi(y-a)/L$ and $U(\xi)=\psi(a+\pi^{-1}L\xi)$. With
\begin{equation}
q(\xi)=\left(\frac{L}{\pi}\right)^2V\left(a+\frac{L}{\pi}\xi\right), \qquad
h=\frac{L}{\pi}\rho,
\end{equation}
the problem \eqref{hgrp} becomes $-U''+qU=\lambda U$ with $U'(0)=hU(0)$ and $U'(\pi)=-hU(\pi)$ where derivatives are taken with respect to $\xi$. Moreover, $\lambda_n=(L/\pi)^2\Lambda_n$ with $\Lambda_n=(\pi/L)^2\lambda_n$. The same rescaling applies to the comparison pair and its spectrum. Furthermore, it preserves integrability and maps reflection about the midpoint of $I$ to $\xi\mapsto\pi-\xi$. The rescaled eigenvalues are strictly ordered and agree with the comparison spectrum for every $n>N$. Lemma~1 in Sec.~5 of \cite{Hald1978} therefore gives a symmetric integrable potential $q$ and a real boundary
parameter $h$ realizing the prescribed spectrum. Corollary~1 in Sec.~2 of the same reference establishes uniqueness of $h$ and of $q$ almost everywhere. Since only finitely many eigenvalues differ, the sequence
$(\lambda_n-\widetilde{\lambda}_n)_{n\geq 0}$ is absolutely
summable. Theorem~1 of \cite{Hald1978} together with the regularity observation following Corollary~2 in Sec.~2, then shows that $q-\widetilde{q}$ agrees almost everywhere with a continuous function. Returning to $I$ gives the claimed existence, uniqueness and continuity of the potential difference. The final assertion follows by adding a continuous representative of $\widetilde{V}$.
\end{proof}
With the free Neumann comparison pair $(\widetilde{V},\widetilde{\rho})=(0,0)$, the theorem applies to the spectra in \eqref{htsg} and
\eqref{rational_one_defect_spectrum}. Last but not least, the arithmetic conditions in Sec.~\ref{spectral_arithmetic} and Corollary~\ref{oorf} determine their oscillator-transfer properties.

\section{First-order expansion on a fixed mode band}
\label{app:finite-band-differentiability}

We now justify the quadratic operator-norm remainders used in Sec.~\ref{fixed_hilbert_perturbations}. The input modes belong to a fixed reference band, while their evolution is measured in the full transported Hilbert space. The following lemma establishes the expansion at the reference profile while allowing the operator domain and zero-mode eigenfunction to vary.
\begin{lemma}\label{lfbpr}
Let us assume that $v,K,\alpha,\beta\in W^{2,\infty}(I)$ are real
and that $v$ and $K$ are uniformly positive. Moreover, $v_\eta,K_\eta$ are given by \eqref{pp} with $|\eta|<\eta_0$ chosen small enough to preserve positivity. By $B_\eta$ and $h_\eta=B_\eta^{1/2}$, we denote the operators in \eqref{fhslo}. 
Let $\mathsf{P}_N$ project onto the first $N$ positive reference modes and set $T_\eta=T_0+\eta\dot{T}$. Then $F_N(\eta):=e^{-ih_\eta T_\eta}\mathsf{P}_N$ is differentiable at $\eta=0$ as a map into
$\Bop(\operatorname{Ran}\mathsf{P}_N,L^2(I))$ and satisfies
\begin{equation}\label{Fbp_exp}
F_N(\eta)=F_N(0)+\eta F_N'(0)+O_N(\eta^2)
\end{equation}
in operator norm. The map $\eta\mapsto\mathcal{R}_\eta^{\mathrm{prof}}\mathsf{P}_N$ is twice continuously differentiable and satisfies the corresponding first-order Taylor estimate. The remainder bounds are uniform when $\dot{T}$ ranges over a bounded set.
\end{lemma}
\begin{proof}
The transported forms have the common domain $H^1(I)$ and depend analytically on $\eta$. After adding the $L^2$ inner product, they are uniformly coercive for sufficiently small $|\eta|$. Form perturbation theory therefore gives analytic resolvents and isolated spectral projections \cite{Kato1995}. The lowest eigenvalue remains exactly zero with normalized eigenfunction $f_{0,\eta}(x)=\sqrt{w_\eta(x)/W_\eta}$ and $W_\eta=\int_I w_\eta(x)~dx$. Its projection is analytic and the positive spectrum is separated from zero for sufficiently small $|\eta|$. We first establish a propagator estimate on the common form domain. The coefficients of the coercive form associated with $B_\eta+\id$ extend holomorphically to a complex neighborhood of $\eta=0$. Resolvent calculus makes $(B_\eta+\id)^{-1/2}$ holomorphic as an operator on $L^2(I)$. The uniform square-root estimates for elliptic forms with
lower-order terms and natural boundary conditions give a locally uniform bound from $L^2(I)$ to $H^1(I)$ (see Theorem~1.1 in \cite{AxelssonKeithMcIntosh2006}). Together, these properties give holomorphic dependence into $H^1(I)$. Inverting this map shows that $(B_\eta+\id)^{1/2}$ depends analytically on $\eta$ in $\Bop(H^1(I),L^2(I))$. The bounded correction $h_\eta-(B_\eta+\id)^{1/2}$ is also analytic. To see this, we separate the zero projection, on which the correction equals $-\id$ and apply resolvent calculus on the positive spectral subspace to $\sqrt{z}-\sqrt{z+1}$, which is holomorphic on a sector containing the positive spectrum and decays at infinity. It follows that
\begin{equation}\label{Lip_fbr}
\|h_\eta-h_0\|_{\Bop(H^1(I),L^2(I))}\leq C|\eta|.
\end{equation}
Since $\operatorname{dom}h_\eta=H^1(I)$, Duhamel's formula applies on this common domain. Moreover, the unperturbed evolution is uniformly bounded in $H^1(I)$ because its graph norm is equivalent to the form norm. Combining \eqref{Lip_fbr} with $|T_\eta-T_0|=|\eta\dot{T}|$ therefore gives
\begin{equation}\label{Lip_fbp}
\left\|e^{-ih_\eta T_\eta}-e^{-ih_0 T_0}\right\|_
{\Bop(H^1(I),L^2(I))}\leq C|\eta|,
\end{equation}
uniformly for bounded $\dot{T}$. We now continue the finite reference band through its Riesz projection $\mathsf{P}_N(\eta)$, with $\mathsf{P}_N(0)=\mathsf{P}_N$. The form resolvent is analytic with values in $H^1(I)$, so the same is true of $\mathsf{P}_N(\eta)$.
Writing $\mathsf{G}_N=\mathsf{P}_N'(0)\mathsf{P}_N$ gives $(\id-\mathsf{P}_N(\eta))\mathsf{P}_N=-\eta\mathsf{G}_N+O_N(\eta^2)$ in $\Bop(\operatorname{Ran}\mathsf{P}_N,H^1(I))$. Let us set $U_\eta=e^{-ih_\eta T_\eta}$. The operator $A_N(\eta)=U_\eta\mathsf{P}_N(\eta)\mathsf{P}_N$ depends analytically on $\eta$ because it involves only the continued finite positive spectral band. If we split the fixed input according to $\mathsf{P}_N(\eta)$ and use \eqref{Lip_fbp} on $\operatorname{Ran}\mathsf{G_N}$, we obtain 
\begin{equation}
F_N(\eta)=A_N(\eta)-\eta U_\eta\mathsf{G}_N+O_N(\eta^2)
=F_N(0)+\eta\left[A_N'(0)-U_0\mathsf{G}_N\right]+O_N(\eta^2).
\end{equation}
This proves \eqref{Fbp_exp}. For the reflection operator, \eqref{tpr} expresses $\mathcal{R}_\eta^{\mathrm{prof}}$ as $M_{r_\eta}\mathsf{p}$. The multiplier $r_\eta$ depends analytically on $\eta$ in $L^\infty(I)$, so Taylor's theorem directly gives
the stated operator-norm remainder.
\end{proof}

\section{Operator realization and spectral asymptotics}
\label{operator_asymptotics}

We prove Propositions~\ref{wto} and \ref{rhea}. More precisely, 
we first identify the weighted operator and its domain from
the quadratic form, then derive the high-energy estimate by
comparison with the free problem at the same Robin parameter.
\begin{proof}[Proof of Proposition~\ref{wto}]
The bounds in \eqref{sr} make the weighted and ordinary
$L^2$ norms equivalent. They also make the form norm
$\left(\mathfrak{a}[u,u]+\|u\|_{\Hil_1}^2\right)^{1/2}$
equivalent to the $H^1(I)$ norm. Since $H^1(I)$ is dense in $\Hil_1$, the form is densely defined, closed and nonnegative. The representation theorem therefore gives a nonnegative self-adjoint operator $A$ characterized by $\mathfrak{a}[u,z]=\langle Au,z\rangle_{\Hil_1}$ for $u\in\operatorname{dom}A$ and every $z\in H^1(I)$. Testing this identity against smooth functions supported
in the interior gives $-(\mathsf{a} u')'=wAu$ in the distributional sense. Hence, $\mathsf{a} u'\in H^1(I)$. Integration by parts with arbitrary endpoint values of $z$ then forces the flux $\mathsf{a} u'$ to vanish at both endpoints. Conversely, every function in \eqref{wtdom} satisfies the weak identity with the differential expression in \eqref{wtop} and this establishes the stated operator domain. Notice that the compact embedding of $H^1(I)$ into $\Hil_1$ gives compact resolvent. Moreover, $\mathfrak{a}[u,u]=0$ precisely when $u$ is constant, so the normalized zero mode is $u_0=W^{-1/2}$ and the eigenvalues are simple by regular Sturm--Liouville theory
with separated boundary conditions~\cite{Zettl2005}. When $v$ and $K$ are even, reflection commutes with $A$. Simplicity therefore gives each eigenfunction a definite parity. Sturm's oscillation theorem gives exactly $n$ simple interior zeros for $u_n$. These zeros occur in reflected pairs  with a midpoint zero
precisely for odd eigenfunctions. Thus, $u_n(-x)=(-1)^n u_n(x)$.
\end{proof}
The high-energy estimate follows by keeping the Robin parameter fixed and comparing the transformed operator with its free counterpart.
\begin{proof}[Proof of Proposition~\ref{rhea}]
The operator $\mathcal L_{\widehat{q},\rho}$ is associated
with the form
\begin{equation}
\mathfrak{l}_{\widehat{q},\rho}[\psi]=v_0^2\int_a^b|\psi'(y)|^2~dy
+\int_a^b\widehat{q}(y)|\psi(y)|^2~dy+v_0^2\rho\left(|\psi(a)|^2+|\psi(b)|^2\right),
\end{equation}
whose domain is $H^1(I)$. The trace inequality and boundedness of $\widehat{q}$ ensure that this form is closed and bounded below for every fixed real $\rho$. Let $\mu_n^{(\rho)}$ denote the eigenvalues of the free operator with $\widehat{q}=0$ and the same Robin parameter, indexed from the lowest eigenvalue by $n=0,1,\ldots$.
The min--max principle gives
\begin{equation}\label{minmax_eb}
\mu_n^{(\rho)}-\|\widehat{q}\|_\infty\leq\lambda_n
\leq\mu_n^{(\rho)}+\|\widehat{q}\|_\infty.
\end{equation}
For the free problem, we introduce $z=y-a$ in $[0,L]$ and write
a positive eigenvalue as $\mu=v_0^2\kappa^2$ with $\kappa>0$.
The solution $\psi_\kappa(z)=\cos{(\kappa z)}+(\rho/\kappa)\sin{(\kappa z)}$ satisfies the left boundary condition.
The right boundary condition gives $(\rho^2-\kappa^2)\sin{(\kappa L)}
+2\rho\kappa\cos{(\kappa L)}=0$. Writing the same solution as a multiple of $\cos{(\kappa z-\theta_\kappa)}$, where $\theta_\kappa=\arctan{(\rho/\kappa)}$, identifies the
high-energy roots through
\begin{equation}
\kappa_n L=n\pi+2\arctan{\left(\frac{\rho}{\kappa_n}\right)}.
\end{equation}
The corresponding eigenfunction has exactly $n$ interior
zeros, and hence Sturm's oscillation theorem identifies $n$ with
its index in the complete ordered spectrum. Notice that this indexing includes any nonpositive eigenvalues of the comparison operator. If we expand the arctangent for fixed $\rho$ as $n\to\infty$, we obtain the free Robin asymptotics~\cite{FultonPruess1994,Zettl2005}
\begin{equation}
\kappa_n=\frac{\pi n}{L}+\frac{2\rho}{\pi n}+O\left(\frac{1}{n^3}\right),\qquad
\mu_n^{(\rho)}=v_0^2\left[\left(\frac{\pi n}{L}\right)^2
+\frac{4\rho}{L}+O\left(\frac{1}{n^2}\right)
\right].
\end{equation}
Combining these expressions with \eqref{minmax_eb} proves \eqref{rla}. In particular, $\lambda_n>0$ for all sufficiently large $n$ and taking the positive square root yields
\begin{equation}
\left|E_n-\frac{\pi v_0 n}{L}\right|=\frac{\left|\lambda_n-(\pi v_0 n/L)^2\right|}{E_n+\pi v_0 n/L}=O\left(\frac{1}{n}\right).
\end{equation}
This proves \eqref{sea}.
\end{proof}

\section{Additional numerical comparison}\label{ANC}
\begin{table}[htp]
\caption{Comparison of discretizations for $r=3$ using the first eight nonzero modes at dimensionless time $3\pi$. FV denotes the cell-centered finite-volume chain and G the nodal piecewise-linear Galerkin scheme with a consistent mass matrix. Here, $J$ counts cells or elements, giving $J$ degrees of freedom for FV and $J+1$ for G. Both errors decrease by approximately a factor of four when $J$ doubles.}
\label{table_dec}
\centering
\small
\begin{tabular}{c|cc|cc}
  $J$ & $R_{\Lambda,8}^{\mathrm{FV}}$ & $R_{\Lambda,8}^{\mathrm G}$
      & $\varepsilon_{8,J}^{(1),\mathrm{FV}}$ & $\varepsilon_{8,J}^{(1),\mathrm G}$\\
  \hline
   128 & $1.783\times10^{-1}$ & $2.231\times10^{-1}$ & $1.050\times10^{-1}$ & $1.312\times10^{-1}$\\
   256 & $4.459\times10^{-2}$ & $5.572\times10^{-2}$ & $2.627\times10^{-2}$ & $3.281\times10^{-2}$\\
   512 & $1.115\times10^{-2}$ & $1.393\times10^{-2}$ & $6.567\times10^{-3}$ & $8.202\times10^{-3}$\\
  1024 & $2.787\times10^{-3}$ & $3.481\times10^{-3}$ & $1.642\times10^{-3}$ & $2.050\times10^{-3}$
\end{tabular}
\end{table}

\bibliography{DB-DD-PRL-GRAF}

\end{document}